\documentclass{emulateapj}
\usepackage{amsmath}
\usepackage{amsfonts} 
\usepackage{appendix}

\usepackage{float}

\shorttitle{Stacked Spectra}
\shortauthors{Mattia Fumagalli et al.}
\slugcomment{Accepted for publication in ApJ}

\begin{document}
\title{Ages of massive galaxies at $0.5 < \lowercase{z} < 2.0$ from 3D-HST rest-frame optical spectroscopy}

\author{Mattia Fumagalli\altaffilmark{1}, Marijn Franx\altaffilmark{1}, Pieter van Dokkum\altaffilmark{2}, Katherine E. Whitaker\altaffilmark{3}\footnotemark[$\dagger$],  Rosalind E. Skelton\altaffilmark{4}, Gabriel Brammer\altaffilmark{5}, Erica Nelson\altaffilmark{2},
 Michael Maseda\altaffilmark{1}, Ivelina Momcheva\altaffilmark{5}, Mariska Kriek\altaffilmark{7}, Ivo Labb{\'e}\altaffilmark{1}, Britt Lundgren\altaffilmark{8},  Hans-Walter Rix\altaffilmark{6}}
\altaffiltext{1}{Leiden Observatory, Leiden University, P.O. Box 9513, 2300 RA Leiden, Netherlands}
\altaffiltext{2}{Department of Astronomy, Yale University, New Haven, CT 06511, USA}
\altaffiltext{3}{Department of Astronomy, University of Massachusetts, Amherst, MA 01003, USA}
\altaffiltext{4}{South African Astronomical Observatory, P.O. Box 9, Observatory 7935, South Africa}
\altaffiltext{5}{Space Telescope Science Institute, 3700 San Martin Drive, Baltimore, MD 21218, USA}
\altaffiltext{6}{Max Planck Institute for Astronomy (MPIA), Königstuhl 17, 69117 Heidelberg, Germany}
\altaffiltext{7}{Astronomy Department, University of California, Berkeley, CA 94720, USA}
\altaffiltext{8}{Department of Astronomy, University of Wisconsin, Madison, WI 53706, USA}
\footnotetext[$\dagger$]{Hubble Fellow}

\begin{abstract}

We present low-resolution near-infrared stacked spectra from the 3D-HST survey up to $z=2.0$ and fit them with commonly used stellar population synthesis models: BC03 (Bruzual \& Charlot, 2003), FSPS10 (Flexible Stellar Population Synthesis, Conroy \& Gunn 2010), and FSPS-C3K (Conroy, Kurucz, Cargile, Castelli, in prep). 
The accuracy of the grism redshifts allows the unambiguous detection of many emission and absorption features, and thus a first systematic exploration of the rest-frame optical spectra of galaxies up to $z=2$.
We select massive galaxies ($\rm log(M_{*} / M_{\odot}) > 10.8$), we divide them into quiescent and star-forming via a rest-frame color-color technique, and 
we median-stack the samples in 3 redshift bins between $z=0.5$ and $z=2.0$. 
We find that stellar population models fit the observations well at wavelengths below $\rm 6500 \AA$ rest-frame, but show systematic residuals at redder wavelengths.
The FSPS-C3K model generally provides the best fits (evaluated with a $\chi^2_{red}$ statistics) for quiescent galaxies, while BC03 performs the best for star-forming galaxies.
The stellar ages of quiescent galaxies implied by the models, assuming solar metallicity, vary from 4 Gyr at $z \sim 0.75$ to 1.5 Gyr at $z \sim 1.75$, with an uncertainty of a factor of 2 caused by the unknown metallicity. On average the stellar ages are half the age of the Universe at these redshifts.
We show that the inferred evolution of ages of quiescent galaxies is in agreement with fundamental plane measurements, assuming an 8 Gyr age for local galaxies. For star-forming galaxies the inferred ages depend strongly on the stellar population model and the shape of the assumed star-formation history. 

\end{abstract}


\section{Introduction}
In recent years, multi-wavelength surveys at high redshift have revealed a significant evolution of galaxies from redshift $z \sim 2$ to the present epoch. 
The emerging picture is based on a few key observations.
First, the star formation rates (SFRs) of galaxies have declined by a factor of 10 in the last 10 billion years. Different observational techniques agree that
this trend is largely independent of mass (Damen et al. 2009, Karim et al. 2011, Fumagalli et al. 2012).
This decline is accompanied by the evolution of the mass function, that once split into star-forming and quiescent population reveals a differential behavior for the two categories: 
while the number of massive star-forming galaxies remains constant or even declines, the number density of massive quiescent galaxies grows by 0.5-1.0 dex from $z\sim$2 (Muzzin et al. 2013, Ilbert et al. 2013).
An immediate consequence is that the quiescent fraction at the massive end becomes increasingly larger at lower redshifts (Bell et al. 2007; Bundy et al. 2006, Brammer et al 2011).
While at lower redshift massive galaxies ($\rm log(M_{*} / M_{\odot}) > 11$) are dominated by a homogenous group of quiescent, red, early-type objects (Djorgovski \& Davis
1987; Blanton et al. 2003; Kauffmann et al. 2003a),
at redshift z$\sim$1 the population shows a large diversity of colors, structural parameters and SFRs (Abraham et al. 2004, van Dokkum et al., 2011).

An additional insight into the assembly history of galaxies is given by their stellar population parameters, namely their age and metallicity.
In the local universe the light-weighted ages and metallicities (both stellar and gaseous) have been shown to correlate tightly with mass
(e.g. Tremonti et al. 2004, Gallazzi et al. 2005, Gallazzi et al. 2006).
While chemical properties of gas in star-forming objects have been traced up to $z \sim 3$ by emission line studies of Lyman break galaxies (e.g. Erb et al. 2006, Moustakas et al. 2011), studies of stellar population parameters at high redshift have been proven challenging, since they require deep spectroscopy in order to trace the rest-frame continuum. 
Recent works by Gallazzi et al. 2014 and Choi et al. 2014 push stellar population analysis to redshifts of $z\sim 0.7$, where the absorption lines commonly used for metallicity and age determinations (Balmer lines, Mg, Na, etc) fall at the edge of optical spectrographs.
At higher redshifts, the optical rest-frame shifts to the infrared, where observations from the ground are notoriously challenging. Determinations of stellar population parameters at $z>1.5$ are limited to a few bright galaxies (Kriek et al 2009, Onodera eta al. 2012, Toft et al. 2012, van de Sande et al. 2012, Bezanson et al. 2013, Onodera et al. 2014, Mendel et al. 2015) or composite spectra (Whitaker et al. 2013)

\begin{figure*}[!t!h]
\centering
\includegraphics[width=18.2cm]{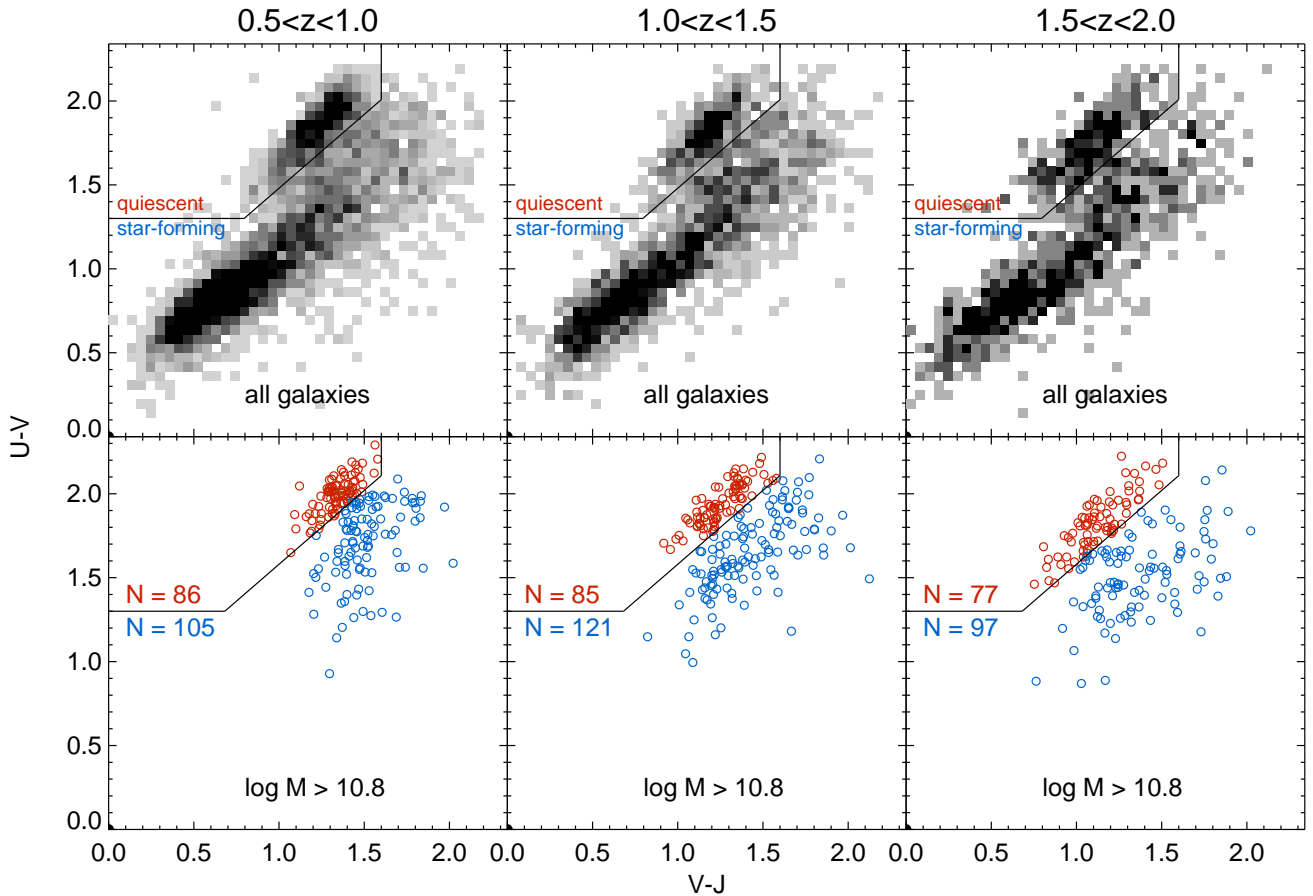}

\caption{Quiescent and star-forming galaxies are separated in a rest-frame color-color diagram. The top row shows, in three redshift bins, galaxies of all masses selected in 3D-HST with the quality cuts described in Section 2.2. Quiescent (red) and star-forming (blue) galaxies which are more massive than $\rm log(M_{*} / M_{\odot}) > 10.8$ are shown in the bottom row.}

\label{UVJplot}
\end{figure*}

In this paper we present observations of galaxies at $0.5 < z < 2.0$ obtained with the low-resolution Wide Field Camera 3 (WFC3) grism onboard Hubble Space Telescope (HST). These spectra cover the observed wavelengths $\rm 11000 < \AA  < 16000$, which correspond to the optical rest-frame for the targeted redshift range. We divide galaxies into quiescent and star-forming, stack their spectra in mass selected samples, and fit them with models from commonly used stellar population synthesis (SPS) codes. 

The goal of the paper is two-fold. In the first place we test the accuracy of SPS models at the observed redshifts and wavelengths. Second, we determine constraints on the stellar ages of galaxies in mass-selected samples, at previously unexplored redshifts.

We note that we apply and test the models in a relatively new regime, both in terms of redshifts and in terms of spectral resolution. Most model tests have been done either at very low spectral resolution (i.e., broad-band and medium-band imaging, with R up to $\sim 8$), or at moderate to high spectral resolution ($R\gtrsim 5000$). Here we apply the models to spectra with $R=50-100$, intermediate between imaging and typical ground-based spectroscopy.

\section{Data}

\subsection{The 3D-HST survey}

The 3D-HST program (van Dokkum et al. 2011; Brammer et al. 2012) is a 625 arcmin$^2$ survey using HST to obtain low-resolution near-IR spectra for a complete and unbiased sample of thousands of galaxies. (Cycles 18 and 19, PI: van Dokkum).
It observes the AEGIS, COSMOS, GOODS-S and UDS fields with the HST/WFC3 G141 grism over 248 orbits, and it incorporates similar, publicly-available data, in the GOODS-N field (GO:11600; PI:Weiner).
These fields coincide with the area covered by CANDELS (Grogin et al. 2011; Koekemoer et al. 2011)
and have a wealth of publicly available imaging data (U band to 24$\mu$m).
The 3D-HST photometric catalogue is described in Skelton et al. (2014), and it constitutes a fundamental step in interpreting the  
spectra that often contain only a single emission line, if any, by providing a photometric redshift prior to the redshift fitting.

The WFC3 grism spectra have been extracted with a custom pipeline, described in Momcheva et al. (2015). Redshifts have been measured via the combined photometric and spectroscopic information using a modified version of the EAZY code (Brammer et al. 2008). 
The precision of redshifts is shown to be $\sigma (\frac{dz}{1+z})= 0.3 \%$
(Brammer et al. 2012, Momcheva et al. 2015; see Kriek et al. 2015 for a comparison based on the new MOSFIRE redshifts from the MOSDEF survey).

Stellar masses have been determined using the FAST code by Kriek et al. (2009), using Bruzual \& Charlot (2003) models, 
and assuming exponentially declining SFHs, solar metallicity, a Chabrier (2003) IMF, and a Calzetti (2000) dust law.

\begin{figure*}[!t]
\centering
\includegraphics[width=8.5cm]{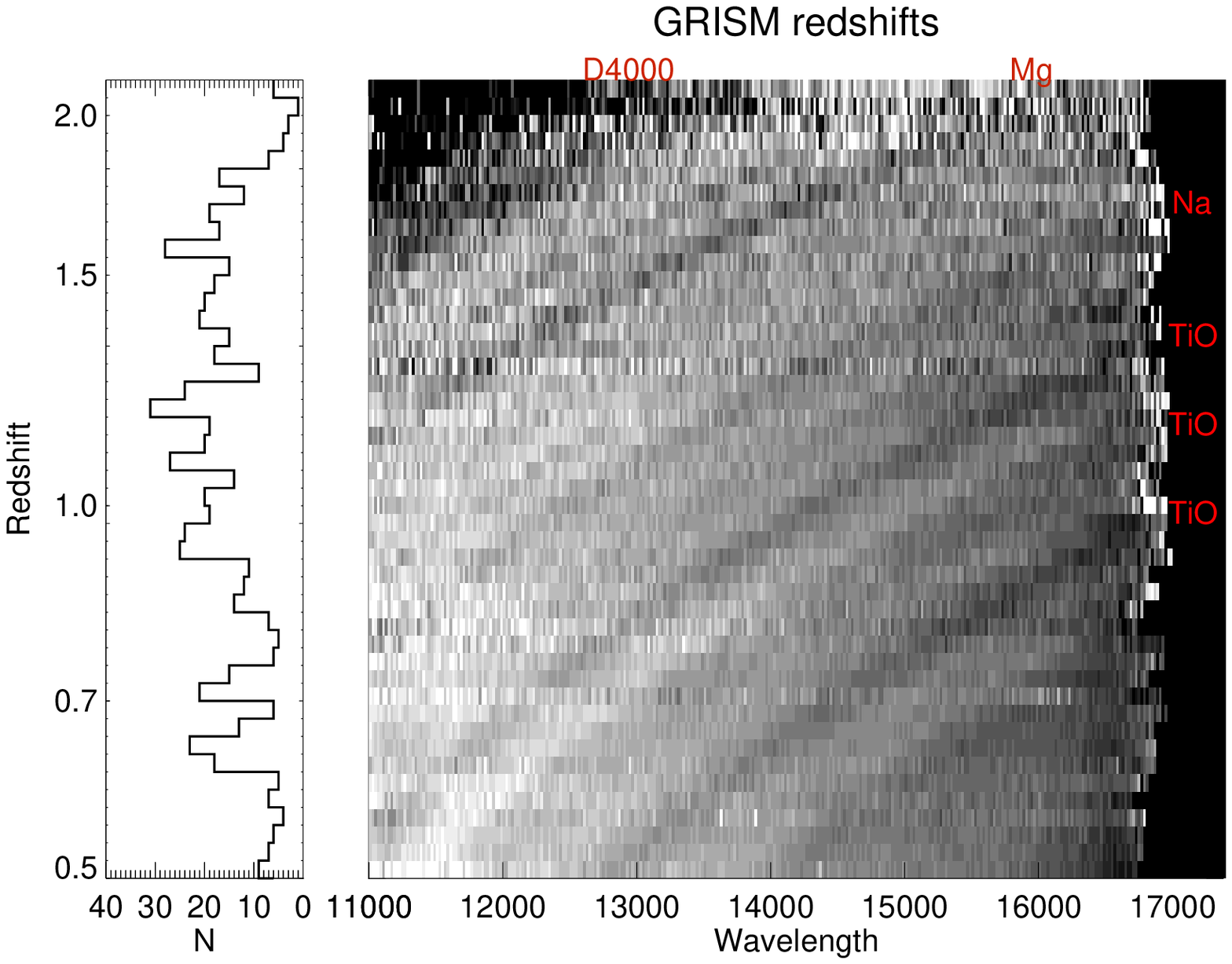} 
\includegraphics[width=8.5cm]{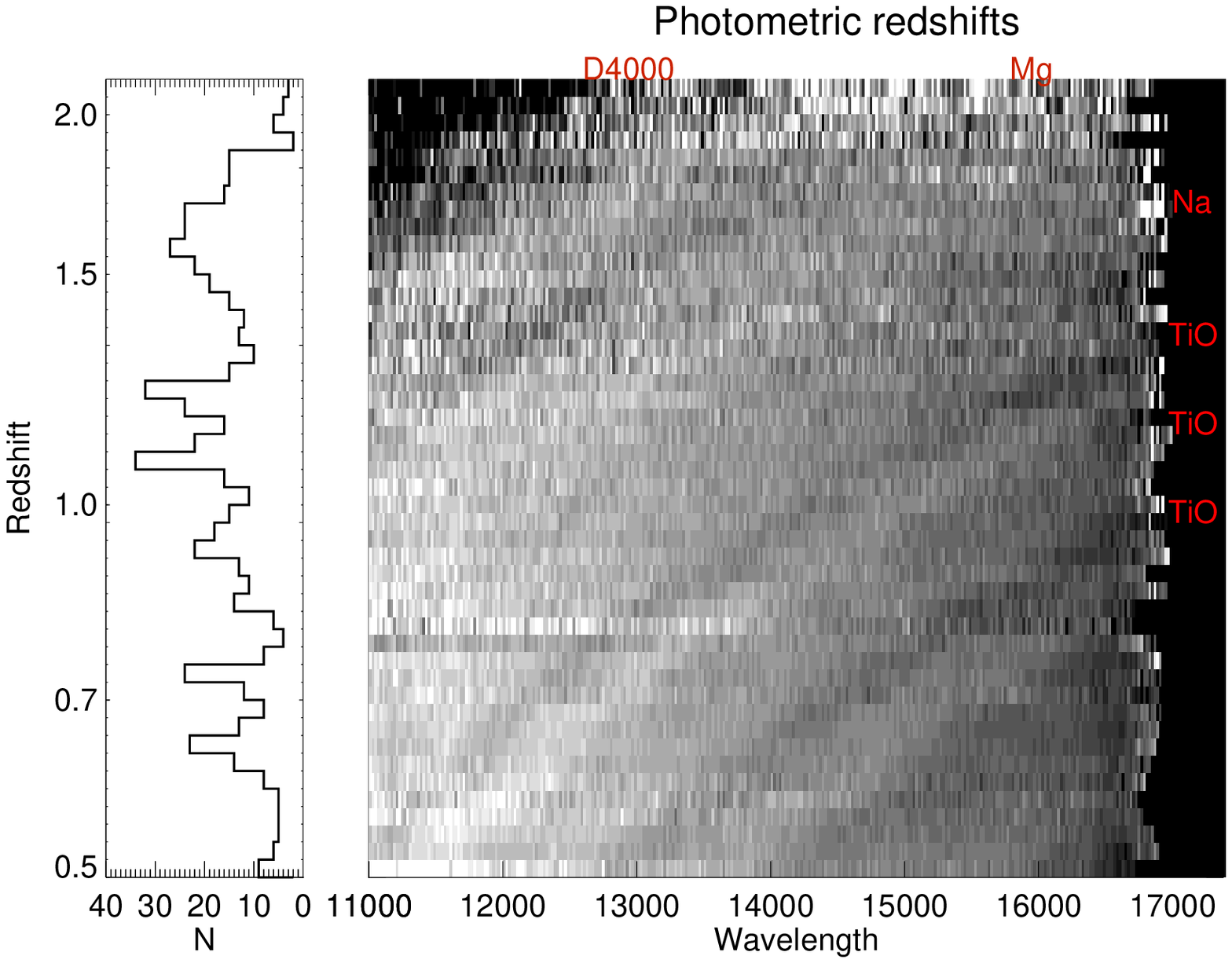}\\

\includegraphics[width=8.5cm]{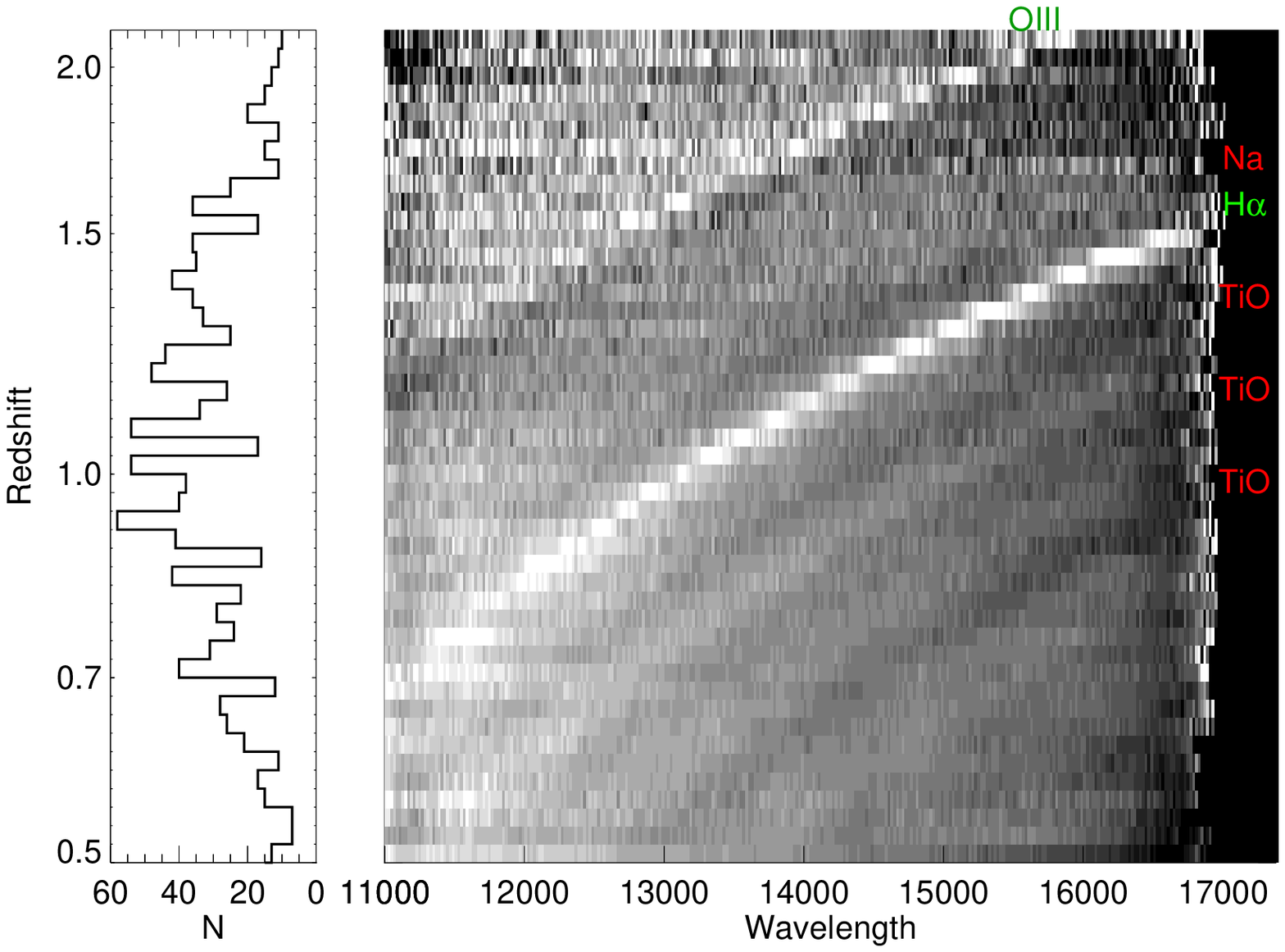}
\includegraphics[width=8.5cm]{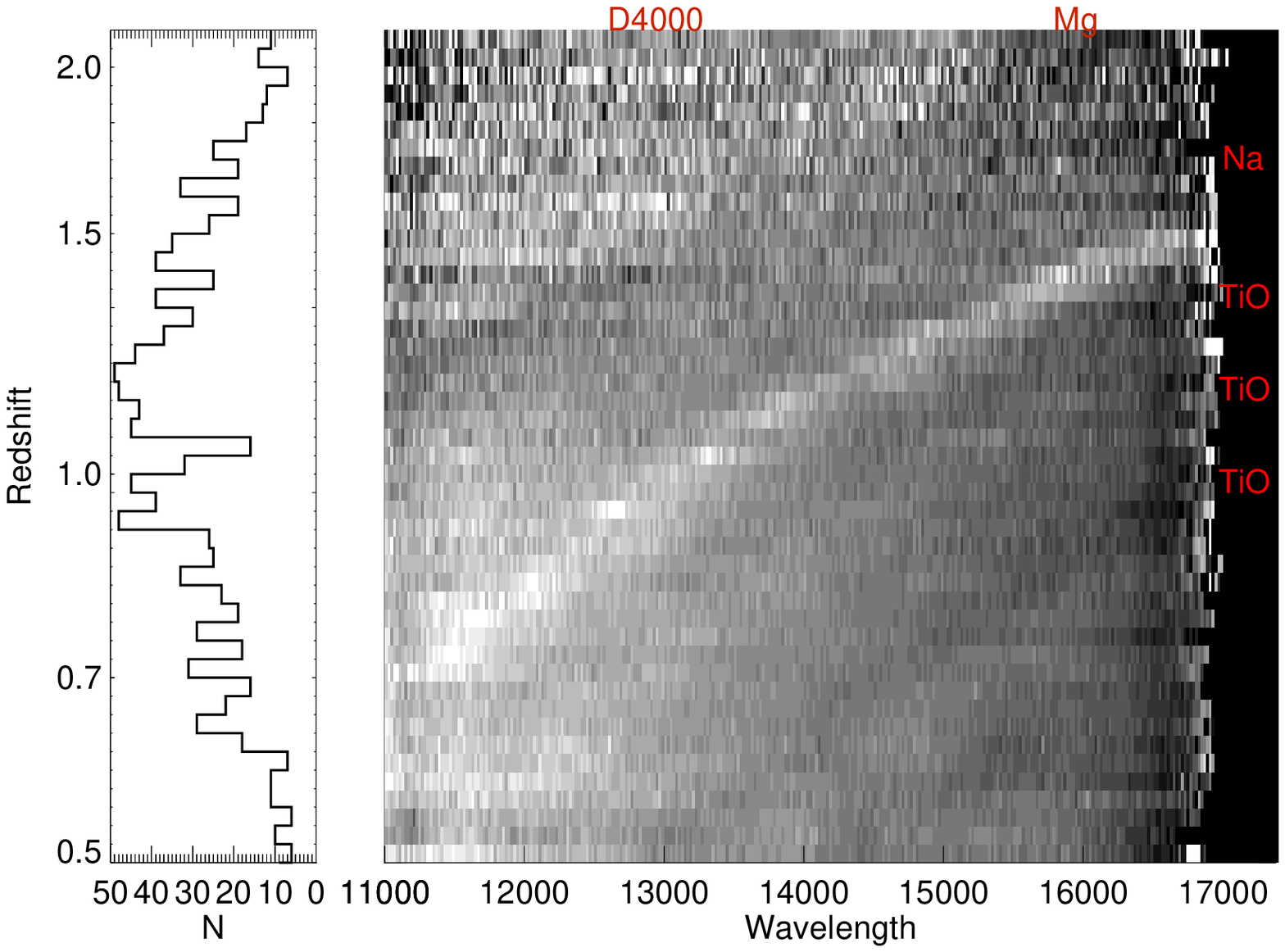}

\caption{Observed spectra of massive ($\rm log(M_{*} / M_{\odot}) > 10.8$) galaxies sorted by redshift, and divided into quiescent (top) and star-forming (bottom).
Galaxies are stacked in a narrow, approximately logarithmic redshift spacing. In the left column grism redshifts are used, while in the right column we take advantage of photometric redshifts only: this demonstrates the quality of grism redshifts and the necessity of high precision in redshift evaluation for stacking galaxies together. The most prominent features in emission and absorption are marked respectively in green and red. No significant emission line is seen in the quiescent sample.
}
\label{ivoplot}
\end{figure*}

\subsection{Sample Selection}

We separate quiescent galaxies from star-forming galaxies using a color-color technique, specifically rest-frame U-V versus rest-frame V-J (hereafter: UVJ diagram).  
It has been noted in the past that selecting quiescent galaxies and star-forming galaxies based on a single color is not reliable, because heavily reddened star-forming galaxies can be as red as quiescent galaxies (among others: Williams et al. 2009).  
Adding information from a second color (V-J) makes it possible to empirically distinguish between galaxies that are red in U-V because of an old stellar population featuring strong Balmer/D4000 breaks (which are relatively blue in V-J) from galaxies that are instead red in U-V because of dust (and therefore are red in V-J too). 

The UVJ diagram has been widely used in a variety of high redshift studies (e.g., Wuyts et al. 2007; Williams et al. 2009; Bell et al. 2012; Gobat et al. 2013), it has been shown to correspond closely to the traditional morphological classes of early-type and late-type galaxies up to at least $z \sim 1$ (Patel et al. 2012) and is able to select dead galaxies with low mid-infrared fluxes (Fumagalli et al. 2014).

Effectively, quiescent galaxies are identified with the criteria $(U-V) > 0.8 \times (V-J) + 0.7$, $U-V > 1.3$ and $V-J < 1.5$ (as in Whitaker et al. 2014). 
The separating lines are chosen with the main criteria being that they lie roughly between the two modes of the population seen in Figure \ref{UVJplot}.

We select galaxies more massive than $\rm log(M_{*} / M_{\odot}) > 10.8$.
In order to achieve a sample of high-quality spectra, we exclude spectra contaminated by neighboring objects for more than 10\% of their total flux, with a wavelength coverage lower than 80\% of the full regime of 1.1 to 1.7 $\mu m$, and with a fraction of bad pixels higher than 10\%. All of these quantities are listed in the 3D-HST catalogs. 
The final sample contains 572 galaxies between redshift 0.5 and 2.0. Figure \ref{UVJplot} shows the selection of massive galaxies, divided into star-forming galaxies and quiescent galaxies, in three redshift bins, superposed on the entire population of galaxies from 3D-HST at the same redshift.

Figure \ref{ivoplot} (left) shows all the spectra in the sample in observed wavelength sorted by redshift and divided into quiescent galaxies (top) and star-forming galaxies (bottom). Spectra are stacked in 50 redshift bins with a roughly exponential spacing. We show the number of galaxies in each bin in the histograms on the right. We see emission and absorption lines being shifted in the wavelength direction, and entering and exiting the observed range at different redshifts. For instance, the H$\alpha$ line enters the wavelength range of the WFC3 grism at $z \sim 0.7$ and exits at $z \sim 1.5$. 
We notice that the subdivision into star-forming galaxies and quiescent galaxies corresponds well to a selection on the presence of emission lines. In the quiescent galaxies sample (Figure \ref{ivoplot}, left) there are no obvious emission lines visible, while at different redshifts we observe deep absorption bands (CaII, Mg, Na, TiO). The star-forming galaxies sample (Figure \ref{ivoplot}, right) features strong emission lines, such as the already mentioned H$\alpha$, and H$\beta$ and [OIII] at higher redshift. Significantly, some absorption bands are detectable also in the SFG sample.
 
 \begin{figure*}[!t!h]
\centering
\includegraphics[width=15.2cm]{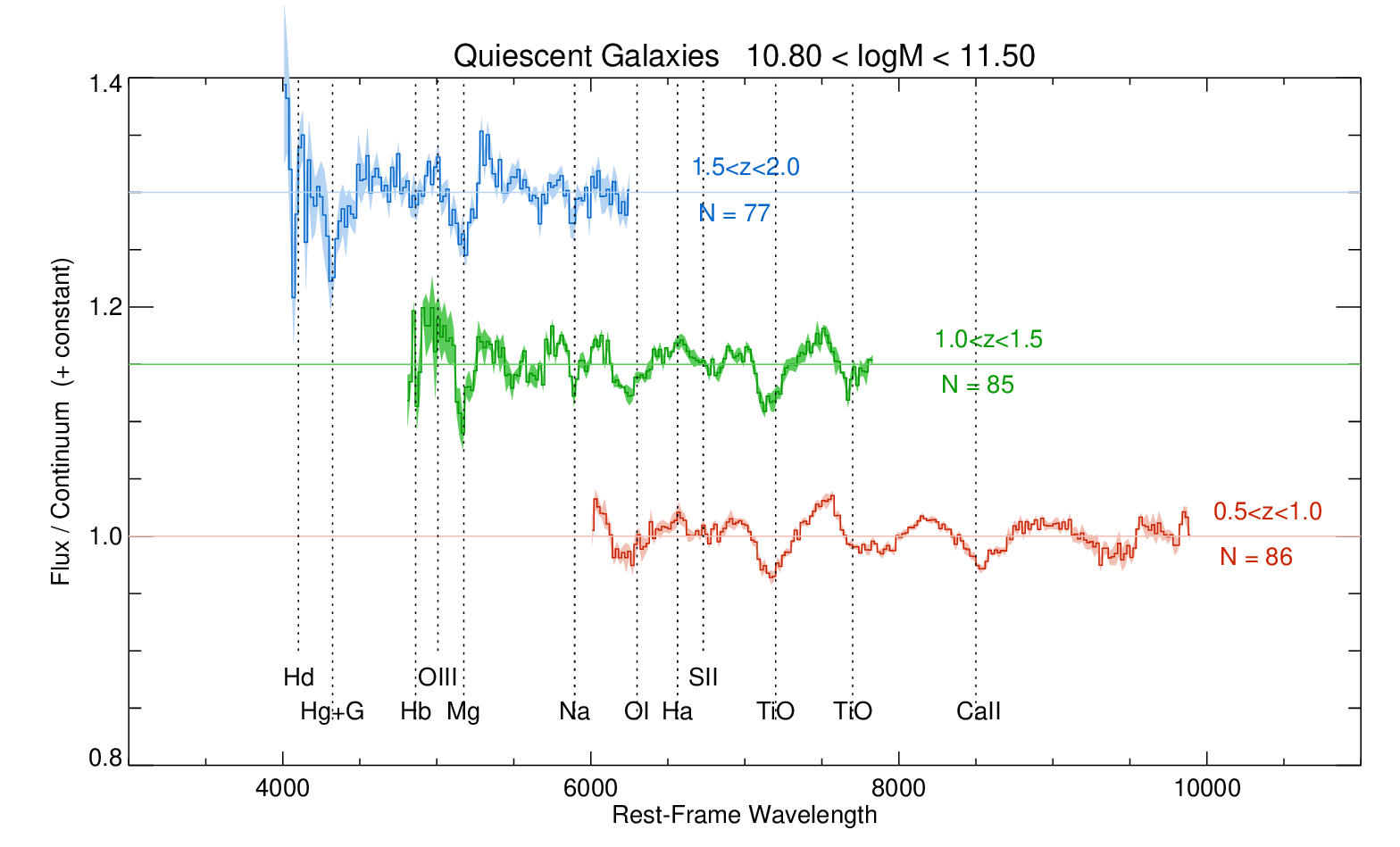}
\caption{Rest-frame stacks of quiescent galaxies with $\rm log(M_{*} / M_{\odot}) > 10.8$, in three redshift bins. The stacks are continuum-subtracted. Shaded regions represent the uncertainty on the stacks derived via bootstrapping (see Section 3.1). Many absorption bands are visible, while no obvious emission lines are seen.}
\label{stacksQG}
\end{figure*}

\begin{figure*}[!t!h]
\centering
\includegraphics[width=15.2cm]{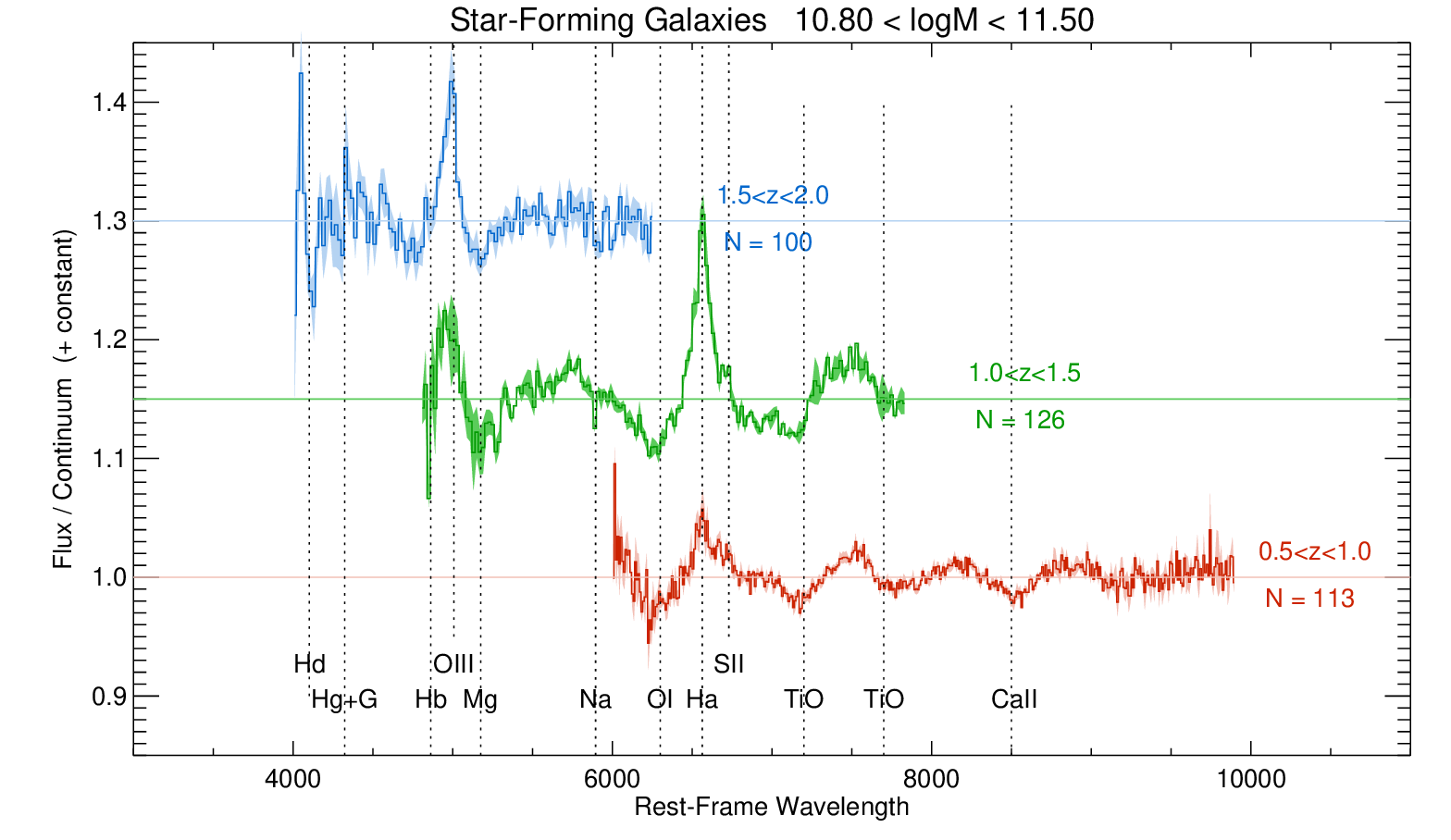}
\caption{Rest-frame stacks of star-forming galaxies with $\rm log(M_{*} / M_{\odot}) > 10.8$, in three redshift bins. The stacks are continuum-subtracted. Shaded regions represent the uncertainty on the stacks derived via bootstrapping (see Section 3.1) Both emission and absorption lines are visible.}
\label{stacksSFG}
\end{figure*}

This experiment proves the quality of 3D-HST grism redshifts. As a comparison, we show in the right column of Figure \ref{ivoplot} the effect of lower quality redshifts on the stacking procedure, by using photometric redshifts instead of grism redshifts. Even though the photometric redshifts provided in the 3D-HST photometric catalogs (Skelton et al. 2014) reach an excellent absolute deviation from spectroscopic redshifts of just $\sigma = 1-2 \%$ (depending on the field), this level of precision is not good enough to clearly observe the spectral features. Figure \ref{ivoplot} shows that the scatter induced by the less accurate photometric redshifts blurs emission and absorption lines. As an example [OIII] is very difficult to track in emission line galaxies if photometric redshifts are used.

This experiment demonstrates that stacking grism spectra requires the sub-percent precision in redshift achieved with the 3D-HST $z_{grism}$.
\section{Methods}
\label{Methods}

\subsection{Stacking}
In individual galaxies in our sample, spectral features are often too weak to be used for reliable measurements of stellar population parameters. 
We therefore achieve the necessary signal-to-noise ratio by stacking spectra in 3 redshift bins, and in the two populations of star-forming galaxies and quiescent galaxies, as follows.
We shift the spectra to rest-frame and fit the continuum in each spectrum with a third order polynomial. In this process we mask regions around known strong emission lines. We normalize the spectra by dividing them by the best-fit polynomial. We next determine the median flux of the normalized rest-frame spectra in a grid of 20$\rm \AA$. 
Errors on the stacks are evaluated via bootstrapping: we perform 100 realizations of each sample by drawing random galaxies from the original sample (repetitions are possible) and we perform the stacking analysis on each resampling. The uncertainty in the flux measurement of each wavelength bin is given by the dispersion of the flux values in the resampled stacks.

The composite spectra are shown in Figure \ref{stacksQG} (for quiescent galaxies) and Figure \ref{stacksSFG} (for star-forming galaxies). Each individual stack is made from the sum of at least 75 galaxies. As the observations cover a constant range of observed wavelength ($\rm 11000 \AA < \lambda < 16000 \AA$), we probe different rest-frame wavelength regimes at different redshifts. The strongest features are the emission lines of H$\alpha$ and [OIII]($\lambda$=5007$\rm \AA$) with a peak strength of 15\% over the normalized continuum. The absorption lines have depths of 5\% or less. The features are weak due to the limited resolution of the spectra.

\subsection{Model fitting}
We compare the stacked spectra with predictions from stellar-population synthesis models (SPS hereafter).
Our goal is to test whether different SPS codes can reproduce the absorption line properties of high redshift galaxies, and to infer stellar ages for these galaxies.
We use models from Bruzual \& Charlot (2003, BC03), Conroy \& Gunn (2010, Flexible Stellar Population Synthesis, FSPS10), and Conroy, Kurucz, Cargile, \& Castelli (in prep, FSPS-C3K)

We use the most standard settings for each SPS code.
The BC03 models are based on the Padova stellar evolution tracks and isochrones (Bertelli et al. 1994); they use the STELIB empirical stellar library
(Le Borgne et al. 2003) for wavelengths between $\rm 3200 \AA < \lambda < 9500 \AA$ and the BaSeL library of theoretical spectra elsewhere.

The FSPS10 models are based on a more updated version of the Padova stellar evolution tracks and isochrones (Marigo et al. 2008); they use the MiLeS empirical stellar library
for wavelengths between $\rm 3500 \AA < \lambda < 7500 \AA$ and the BaSeL library of theoretical spectra library elsewhere. 

We have also considered a new, high resolution theoretical spectral library (FSPS-C3K, Conroy, Kurucz, Cargile, Castelli, in prep).  This library is based on the Kurucz suite stellar atmosphere and spectral synthesis routes (ATLAS12 and SYNTHE) and the latest set of atomic and molecular line lists.  The line lists include both lab and predicted lines, the latter being particularly important for accurately modeling the broadband SED shape.  The grid was computed assuming the Asplund (2009) solar abundance scale and a constant microturbulent velocity of 2 km/s. 

\begin{figure}[!h!]
\centering

\includegraphics[width=8.5cm]{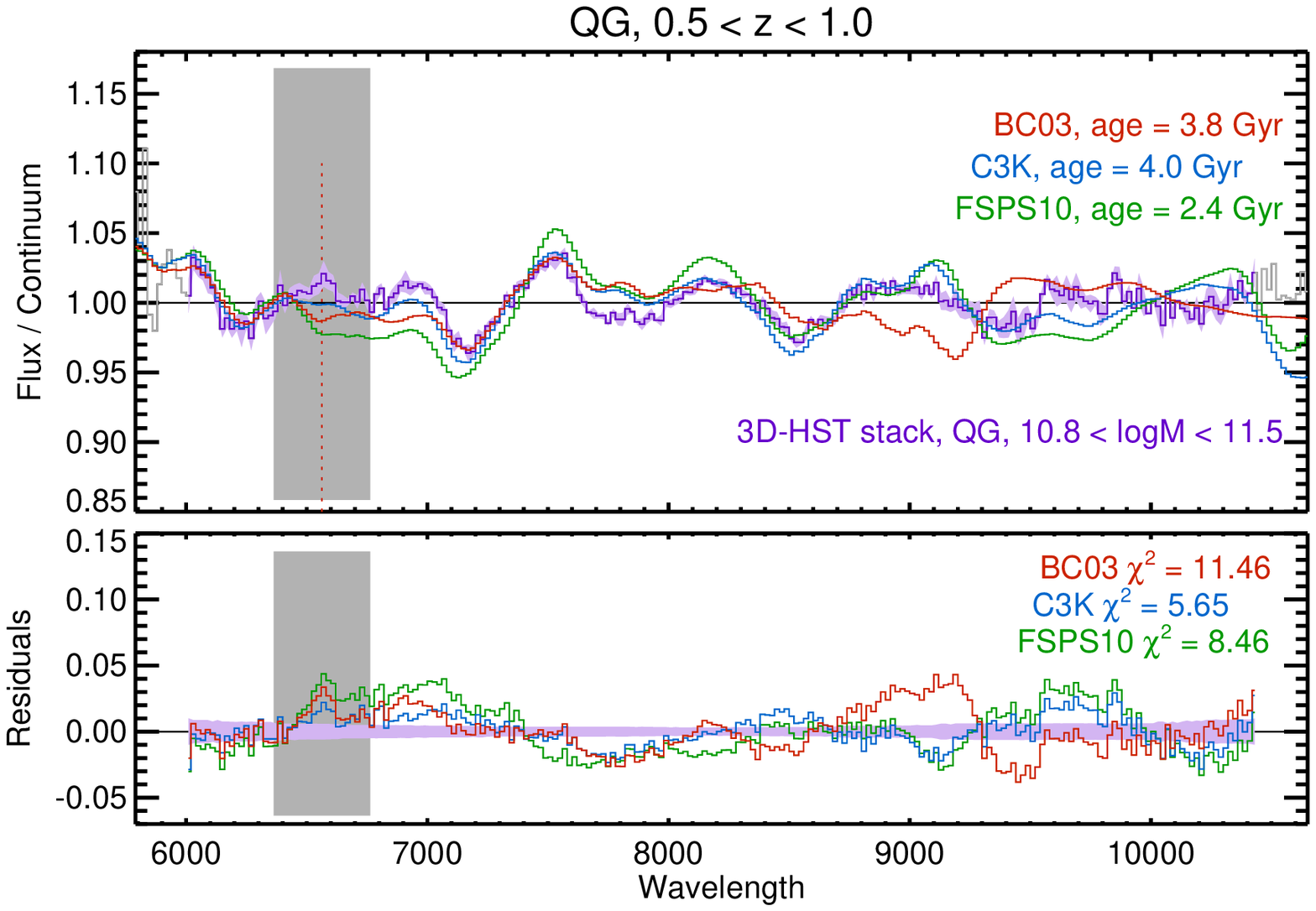}

\caption{Best fits to the stack of quiescent galaxies at 0.5$<z<$1.0, $\rm log(M_{*} / M_{\odot}) > 10.8$ (purple), with BC03 (red), FSPS10 (green), and FSPS-C3K (blue) SSPs. Errors on stacks are computed through bootstrapping on the sample. The grey area represents the wavelength region around H$\alpha$ masked in the fitting process. A comparison of residuals from different models is shown in the bottom panel, where the purple shaded region represents the uncertainty in the stacked spectrum. The FSPS-C3K models provide the lowest $\chi_{red}^2$; its best-fit age is 4.0 Gyr.}
\label{bestfitsQG05}
\end{figure}

\begin{figure}[!h!]
\centering
\includegraphics[width=8.5cm]{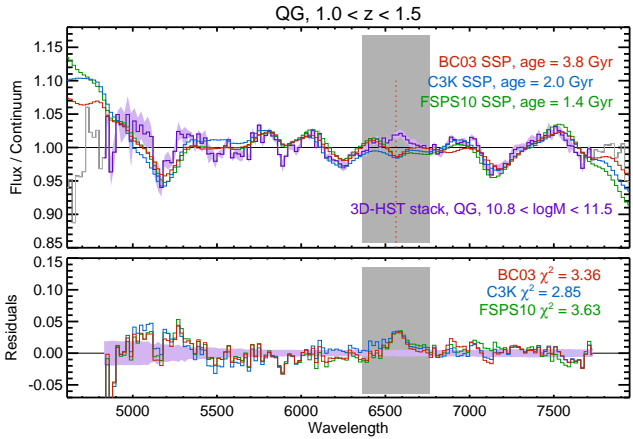}
\caption{Best fits to the stack of quiescent galaxies at 1.0$<z<$1.5, $\rm log(M_{*} / M_{\odot}) > 10.8$, with the same color coding as Figure 5. The $\chi_{red}^2$ of different models are comparable. The age determinations span a wide range from 1.4 to 3.8 Gyr, according to the model in use.}
\label{bestfitsQG10}
\end{figure}

\begin{figure}[!h!]
\centering  
\includegraphics[width=8.5cm]{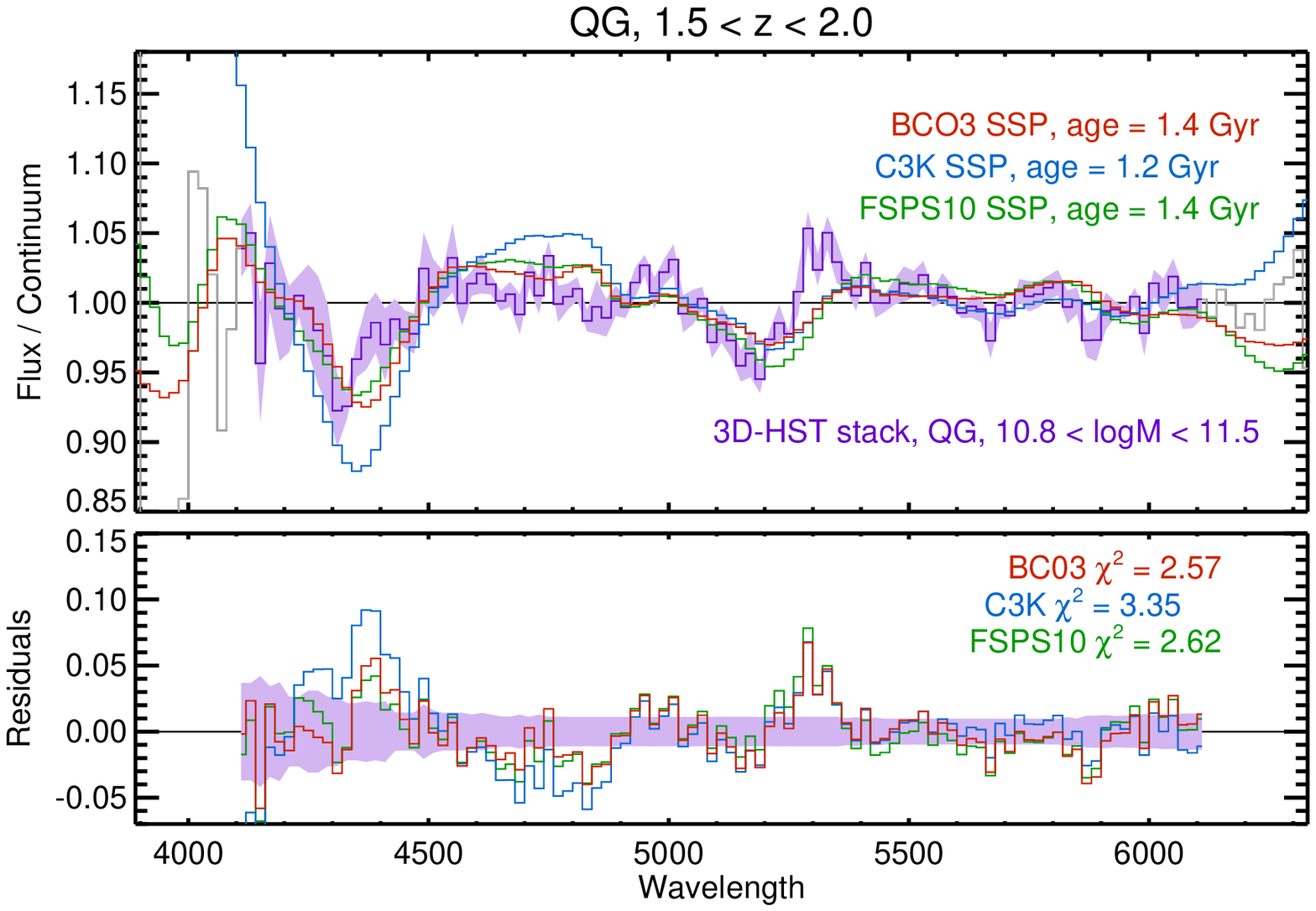}
\caption{Best fits to the stack of quiescent galaxies at 1.5$<z<$2.0, $\rm log(M_{*} / M_{\odot}) > 10.8$, with the same color coding as Figure 5. A comparison of residuals from different models is shown in the bottom panel. Residuals from different models are comparable, and the age determination converges to values of 1.2-1.4 Gyr.}
\label{bestfitsQG15}
\end{figure}

For each observed stack, we perform a least-squares minimization using the three different models, to find the best-fit age of the stack. 
In order to compare the high resolution models with the low resolution stacks, we need to downgrade the models to 3D-HST resolution.
The dispersion of the G141 grism is $\rm 46 \AA$/pixel (R $\sim $130 in the raw data) with a raw pixel scale of 0.12 arcsec, sampled with 0.06 arsec pixels.
The lack of a slit, combined with the low spectral resolution of the WFC3 grism, implies that emission and absorption “lines” are effectively images of the galaxy at that particular wavelength (see Nelson et al. 2015 for a detailed explanation). In a continuum spectrum, this means that the spectral resolution is different for each galaxy, as it is determined by its size.  This effect is generally referred to as "morphological broadening" in the slitless spectroscopy literature (e.g. van Dokkum et al. 2011, Whitaker et al. 2013, 2014).
We simulate the expected morphological broadening by convolving the high-resolution model spectra with the object morphology in the $\rm H_{F140W}$ continuum image collapsed in the spatial direction. Model spectra for each galaxy in the sample are continuum-divided and stacked with the same procedure we use for observed spectra. In the fitting of models to data, we allow an additional 3rd order polynomial continuum component with free parameters. 

In summary, for each sample of galaxies, we create mock 3D-HST stacks based on three stellar population models (BC03, FSPS10, FSPS-C3K) with solar metallicity, Chabrier IMF, and three different star-formation histories (single stellar burst, constant star formation, and an exponentially declining model with $\tau$ = 1 Gyr), with a spacing of 0.1 Gyr in light-weighted age. Unless specified otherwise, all ages quoted through the paper are light-weighted ages.  For an SSP, this age is identical to the time elapsed since the beginning (and end) of star formation, as all stars have the same age. In other models this age is some time in between the time elapsed since the onset of star formation and the ages of the youngest stars (see, e.g., van Dokkum et al. 1998).

Our choice of performing the stacking and the fitting analysis on continuum-divided spectra is motivated by the goal of measuring ages of galaxies without being influenced by the slope of the continuum, which is degenerate in dust and age. Aging stellar populations have redder broadband colors, however dust reddening and increasing metallicity have a similar effect. For instance, the difference in $g-r$ color corresponding to 1 Gyr of passive aging within the BC03 models, can also be caused by 0.5 mag of dust reddening following the Calzetti et al. (2000) dust law, or by an increase in metallicity from log(Z) = 0.02 (solar) to log(Z) = 0.05.

In this study we do not treat galaxies hosting an active galactic nucleus (AGN) separately. We test the influence of AGNs by selecting all sources falling in the IRAC color-color selection presented in Donley et al. (2012). The IRAC selected AGNs count for less than 5$\%$ of each sample of quiescent galaxies/star-forming galaxies at different redshifts. The conclusions of the paper do not change when these sources are removed from the stacks.

\section{Quiescent Galaxies}
\label{Quiescent Galaxies}

We fit the stacks of quiescent galaxies with SSPs from three SPS models (BC03, FSPS10, FSPS-C3K) assuming solar metallicity, a Chabrier IMF, and a single-burst SFH \footnote{We redo the analysis assuming instead an exponentially declining SFH ($\tau$=1 Gyr), finding that light-weighted ages and residuals to the best fits are compatible (within 1$\sigma$) to those measured from the SSP models.}.

In this Section we will discuss separately the quality of the fits for different SPS models, and the stellar ages determined from the best fits.

\subsection{Quality of fits}
\label{Quality of fits QG}

Figure \ref{bestfitsQG05} shows for each of the models (BC03, FSPS10, FSPS-C3K) the best fits to the QG stack for the lower redshift bin ($0.5<z<1.0$). 
The grey shaded area represents the area around H$\alpha$, masked in the fitting. With BC03 (Figure \ref{bestfitsQG05}, red) the best fit is very poor at wavelengths higher than 7500 $\rm \AA$, as shown by the residuals in the lower panel. Moreover, the reduced $\chi^2$ ($\chi_{red}^2$) of the best fit is high (11.46).
Using FSPS10 (green), the best fit also has significant ($> 3\%$) residuals at the reddest wavelengths ($\rm >8000 \AA$) and around the  $\rm > 7000 \AA$ regime, where the first TiO band lies. The $\chi_{red}^2$ value of the best fit is still high (8.5). Finally, using the latest FSPS-C3K models (Figure \ref{bestfitsQG05}, bottom left) the best fit converges with a lower $\chi_{red}^2 = 5.8$, and residuals are below $2\%$ consistently over the entire wavelength range.
We compare residuals from different SPS models (data-model) in the bottom panel of Figure \ref{bestfitsQG05}. All best-fits have positive residuals in the $\rm 7000 \AA$
region (up to $4\%$ for FSPS10), underestimating the fluxes at those wavelengths. At wavelengths higher than 8000$\rm \AA$ the BC03 models have positive residuals, while FSPS10/FSPS-C3K have negative ones. 

Moving to higher redshifts, the quality of fits with different SPS models is comparable. In the intermediate ($1.0<z<1.5$, Figure \ref{bestfitsQG10}) and in the high redshift bin ($1.5<z<2.0$, Figure \ref{bestfitsQG15})  we examine, all $\chi_{red}^2$ range from 2.8 to 3.3. Residuals in these redshift bins are comparable among different models, and tend to be smaller than a few percent. We evaluate residuals corresponding to the H$\alpha$ line in quiescent galaxies in Section \ref{Halpha in quiescent galaxies}.

\subsection{Determination of Ages}
\label{Determination of Ages QG}

The stellar ages of quiescent galaxies implied by the best-fit models vary according to the used SSP.
In order to evaluate the uncertainty of the age measurement we bootstrap the sample 100 times and repeat the fitting analysis on the bootstrapped realizations of the stack.

At the lowest redshift bin ($0.5<z<1.0$, Figure \ref{bestfitsQG05}) 
we obtain a stellar age of 3.8 $\pm 0.6$ Gyr with BC03, a younger age (2.4 $\pm 0.4$ Gyr) with FSPS10 and again 4.0 $\pm 0.2$ Gyr with FSPS-C3K (which is the model with the lowest residuals).
A similar wide range of age determinations is obtained for the intermediate redshift bin (Figure \ref{bestfitsQG10}) ranging from 1.4 $\pm 0.1$ Gyr for FSPS10, to  2.0 $\pm 0.3$ Gyr for FSPS-C3K to 3.8 $\pm 0.8$ Gyr for BC03. 

In the highest redshift bin ($1.5<z<2.0$, Figure \ref{bestfitsQG15}) all the age determinations are between 1.2 and 1.4 Gyr. This value is consistent with Whitaker et al. (2013), who studied a sample of galaxies at slightly different masses and redshifts ($10.3 <\rm log(M_{*} / M_{\odot}) < 11.5$, $1.4 < z < 2.2$) obtaining an age of 1.25 Gyr computed with the Vakzdekis models. We also agree with Mendel et al. (2015), who investigate the stellar population of 25 massive galaxies with VLT-KMOS, deriving a mean age of 1.08 $^{0.13}_{-0.08}$ Gyr.

Our study relies on the assumption that quiescent galaxies already have a solar metallicity at high redshift.
This assumption is supported by the study of Gallazzi et al. (2014), who studied 40 quiescent galaxies at $0.65<z<0.75$ with IMACS spectra, obtaining a mass-metallicity relation consistent with that at $z=0$ from SDSS.

\begin{figure}[!h!]
\centering
\includegraphics[width=8.8cm]{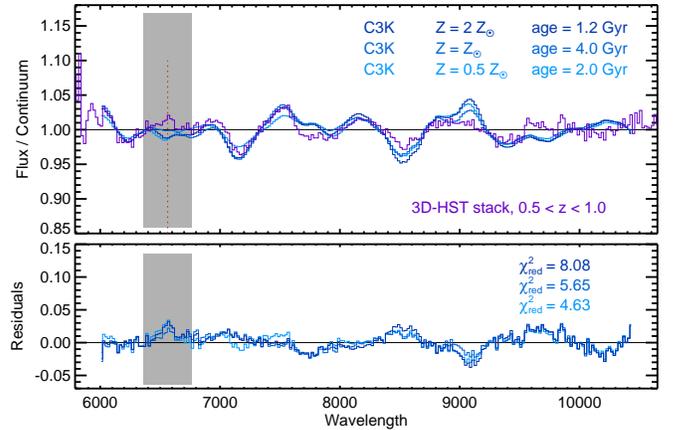}

\caption{Best fits to the stack of quiescent galaxies at 0.5$<z<$1.0, $\rm log(M_{*} / M_{\odot}) > 10.8$, with the FSPS-C3K models and varying metallicity. A comparison of residuals from different models is shown in the bottom panel.
}
\label{bestfitmet}
\end{figure}

We explore the effect of changing metallicity in Figure \ref{bestfitmet} on the stack with the highest signal-to-noise ratio. We use the stellar population model that gave the lowest $\chi_{red}^2$ with the standard solar metallicity (FSPS-C3K), and vary the metallicity to twice solar and half solar. The best fit with a $Z=0.5Z_{\sun}$ metallicity has a $\chi^2_{red}$ marginally lower than that with a solar metallicity (4.6 versus 5.6), while the $Z=2Z_{\sun}$ case can be excluded by its higher $\chi^2_{red}$=8.1. 
We moreover remark that, as seen in Section 4.1, none of the models that we explore properly fit the data at rest-frame wavelengths redder than ~7500$\rm \AA$: at those wavelengths lines such as TiO and CaII could help in constraining metallicities  (e.g. Carrera et al. 2007, Conroy et al. 2014), but improved models or higher resolution spectroscopy are required. 

At higher redshifts we find that the uncertainty on the ages due to the uncertainty in the metallicity can amount to a factor of 2, with the $Z=2Z_{\sun}$ and $Z=0.5Z_{\sun}$ best fits being respectively younger and older than those with a solar metallicity, as expected from the well-known degeneracies between age and metallicity.

This metallicity uncertainty dominates the error budget.

\begin{figure*}[!t!h]
\centering
\includegraphics[width=8.8cm]{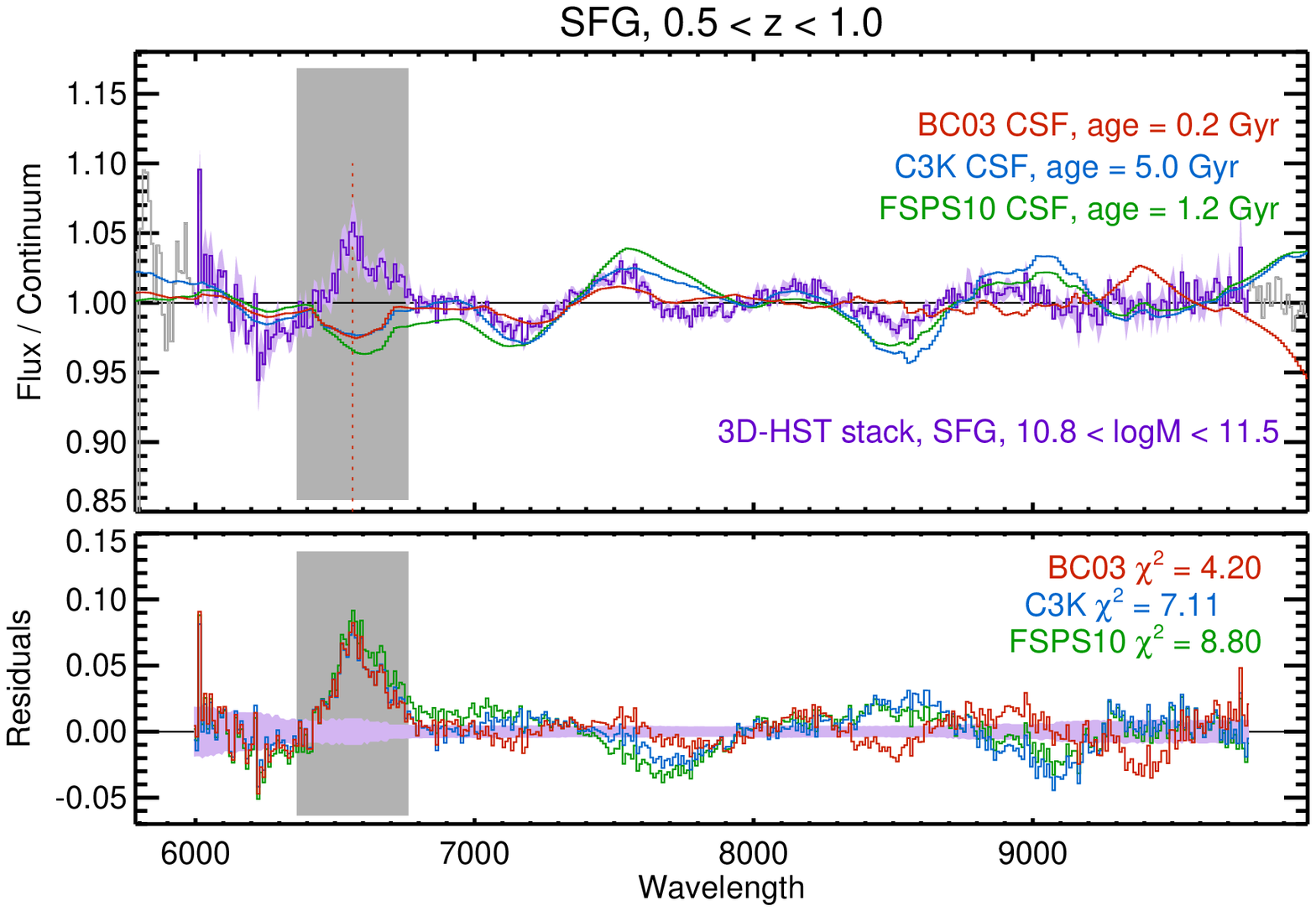}
\includegraphics[width=8.8cm]{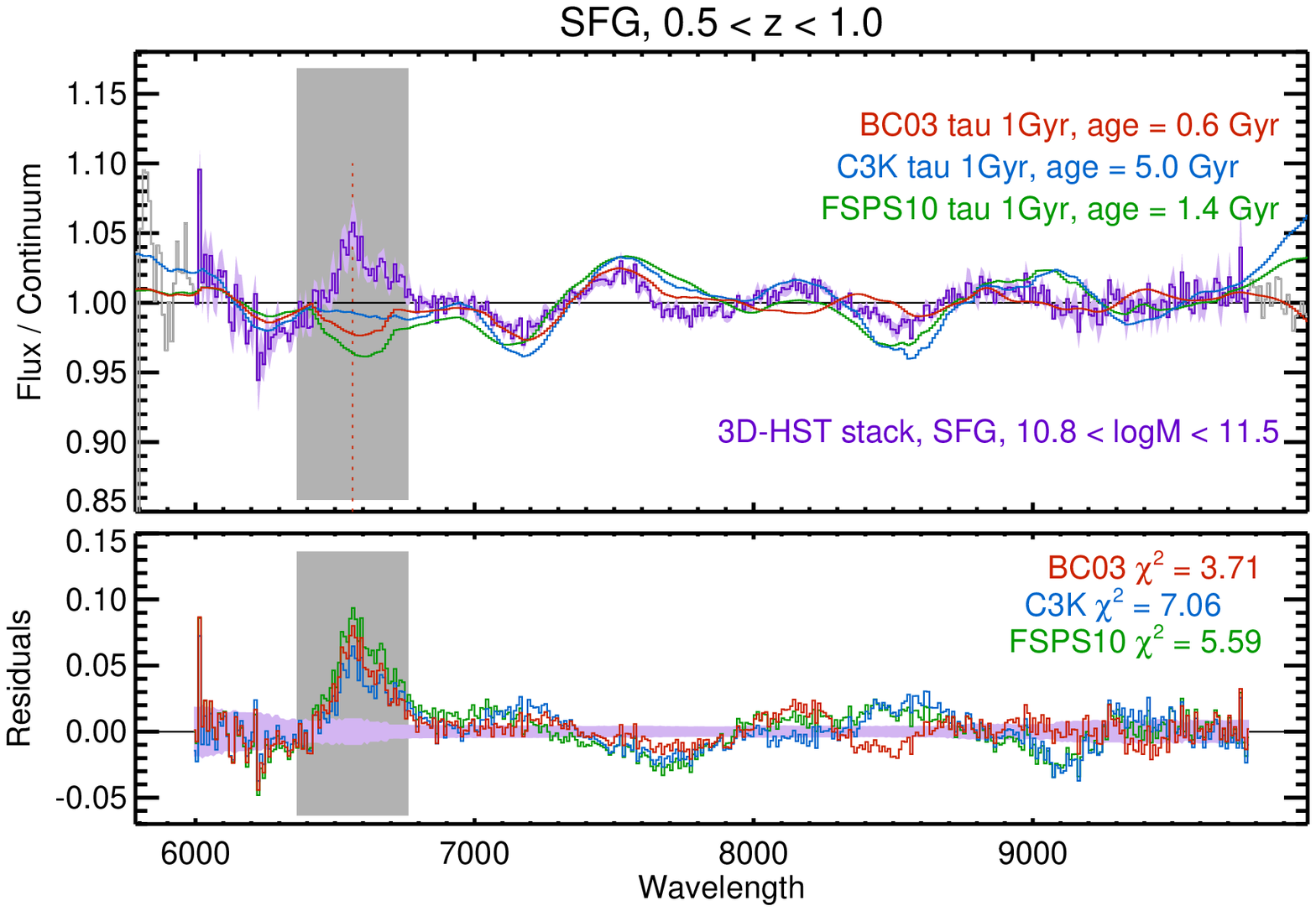}\\
\caption{Best fits to the stack of star-forming galaxies at 0.5$<z<$1.0, $\rm log(M_{*} / M_{\odot}) > 10.8$, with BC03 (red), FSPS10 (green) and FSPS-C3K (blue) models. The grey area represents the wavelength region around H$\alpha$ masked in the fitting process. On the left models with a constant star formation rate are used; on the right, exponentially declining models with $\tau$ = 1 Gyr. BC03 models provide the best fits, according to a $\chi_{red}^2$ statistic. Best-fit ages vary significantely among SPS models and assumed SFHs.}
\label{stacksSFG05}
\end{figure*}

\begin{figure*}[!t!h]
\centering
\includegraphics[width=8.8cm]{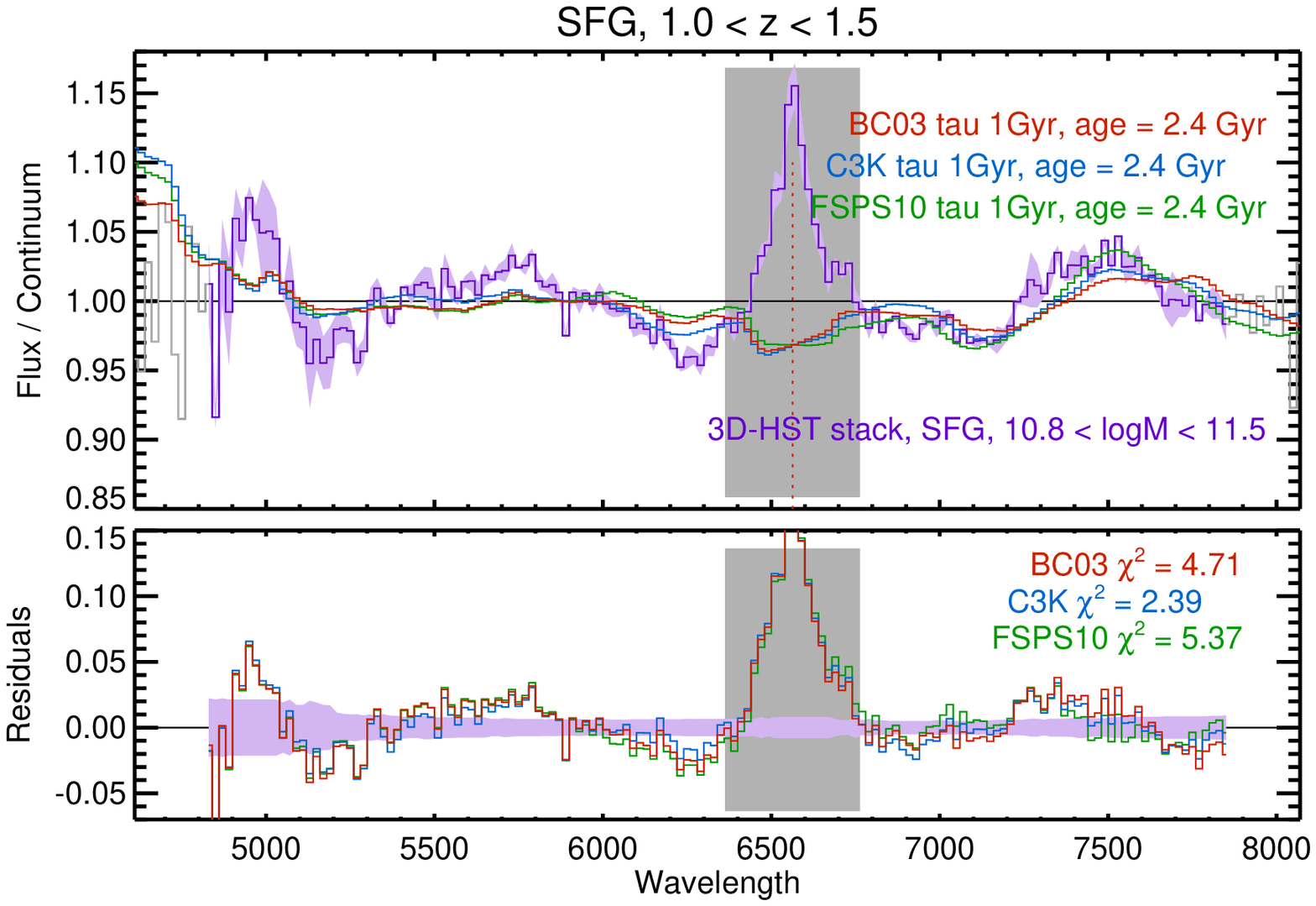}
\includegraphics[width=8.8cm]{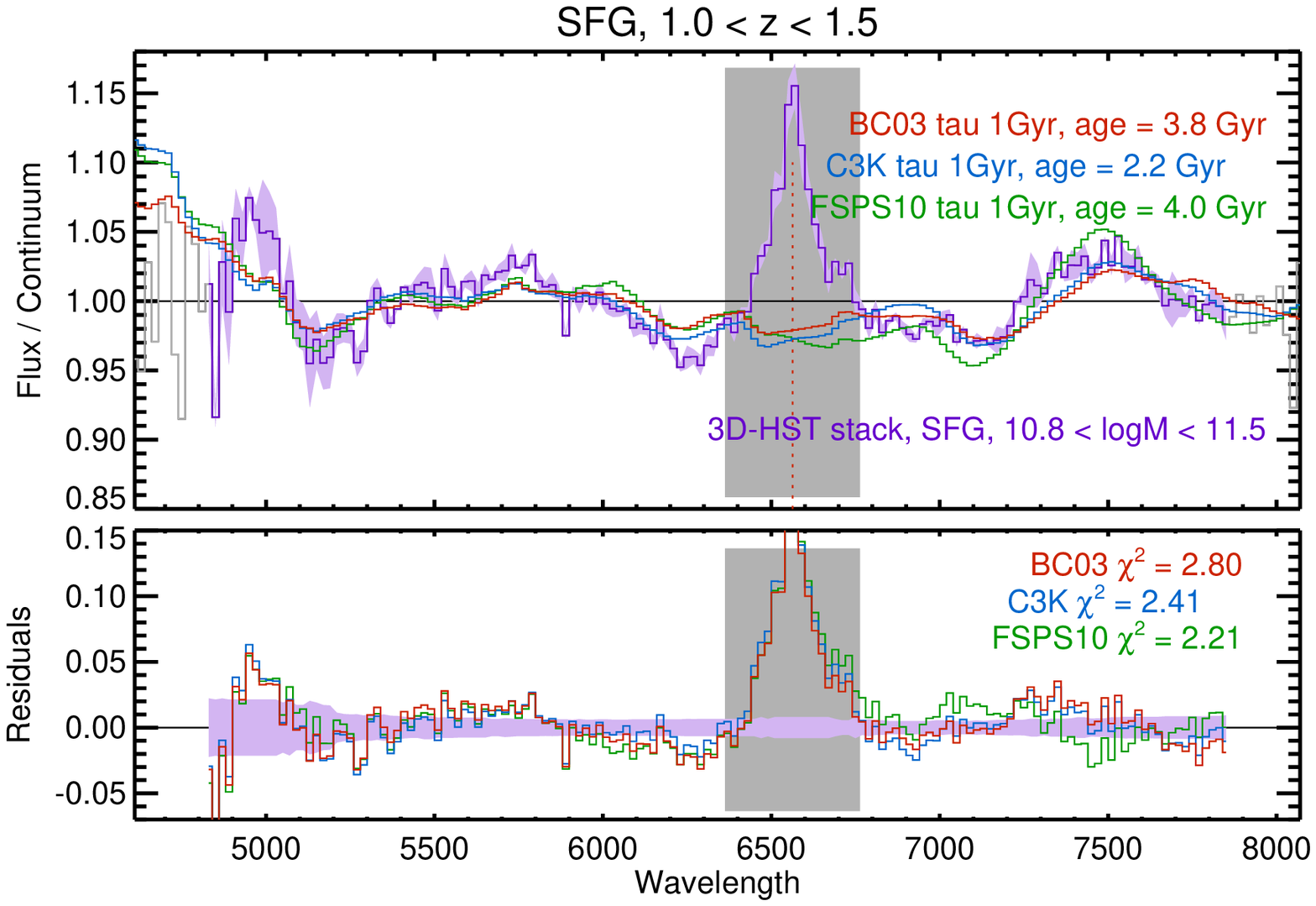}
\caption{Best fits to the stack of star-forming galaxies at 1.0$<z<$1.5, $\rm log(M_{*} / M_{\odot}) > 10.8$, with BC03 (red), FSPS10 (green) and FSPS-C3K (blue) models. The grey area represents the wavelength region around [OIII] masked in the fitting process. On the left models with a constant star formation rate are used; on the right, exponentially declining models with $\tau$ = 1 Gyr. Best fit ages vary from 2 to 4 Gyr.}
\label{stacksSFG10}
\end{figure*}

\begin{figure*}[!t!h]
\centering
\includegraphics[width=8.8cm]{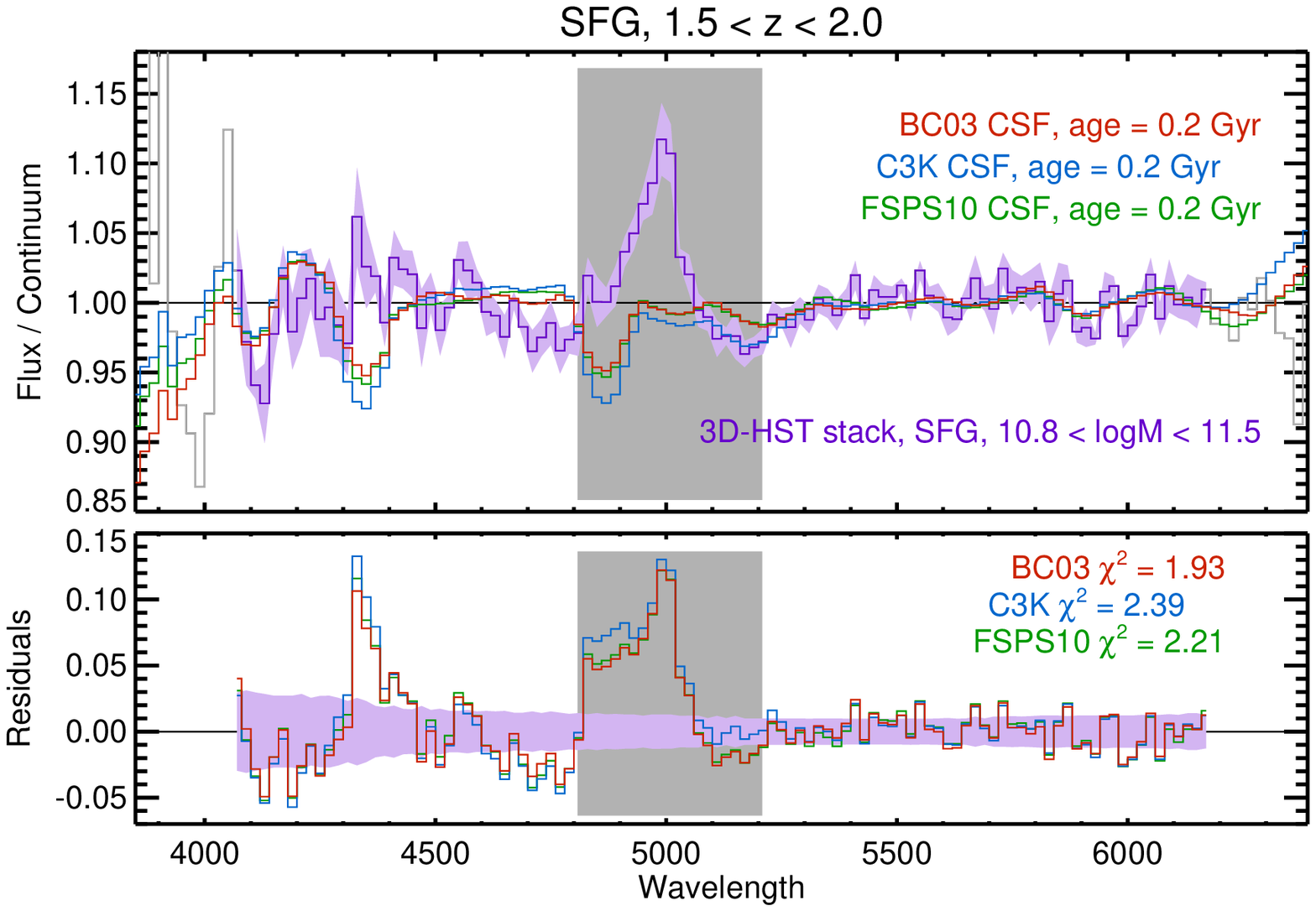}
\includegraphics[width=8.8cm]{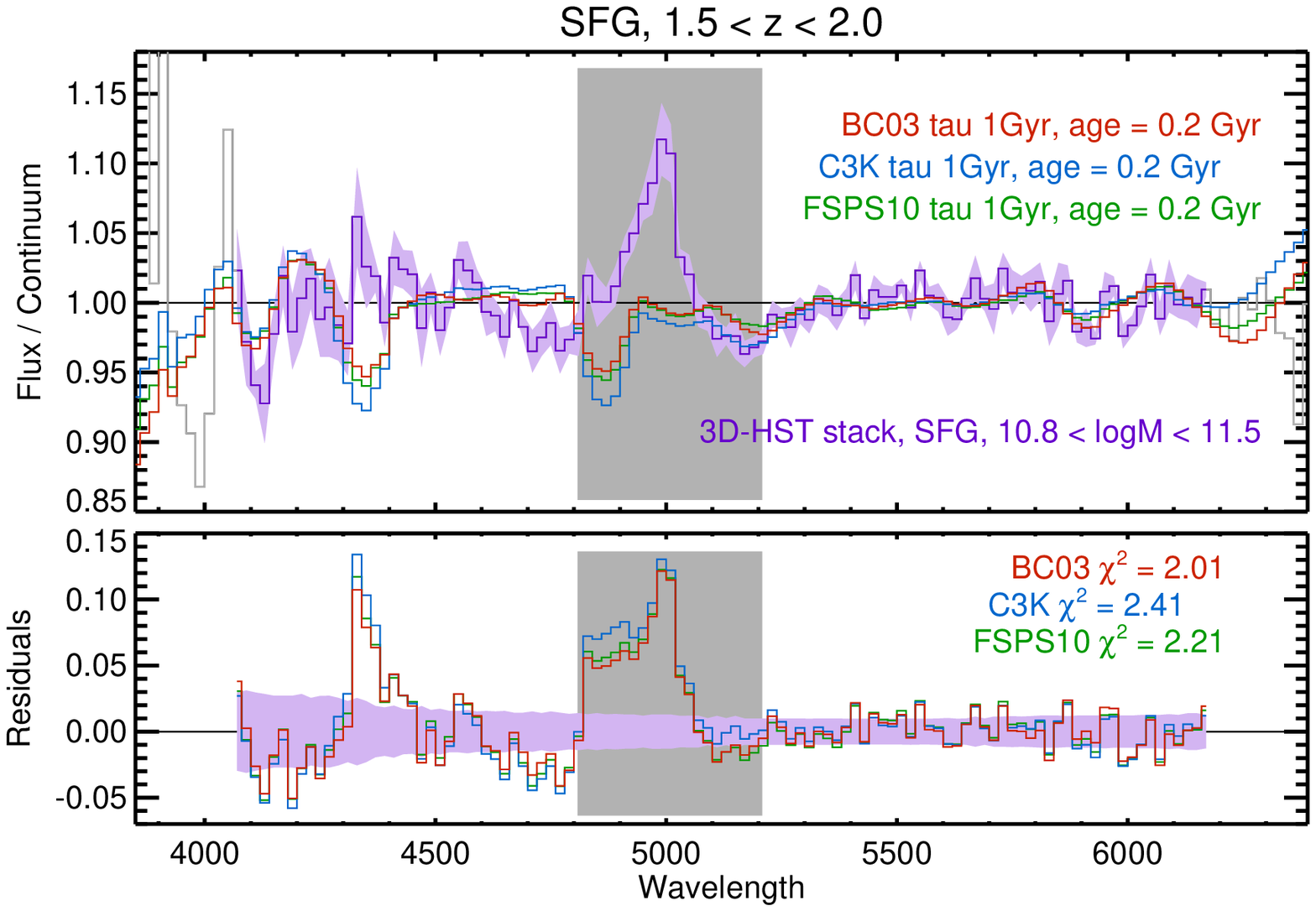}
\caption{Best fits to the stack of star-forming galaxies at 1.5$<z<$2.0, $\rm log(M_{*} / M_{\odot}) > 10.8$, with BC03 (red), FSPS10 (green) and FSPS-C3K (blue) models. The grey area represents the wavelength region around [OIII] masked in the fitting process. On the left models with a constant star formation rate are used; on the right, exponentially declining models with $\tau$ = 1 Gyr. Different SPS models have similar residuals and converge to very young ages.}
\label{stacksSFG15}
\end{figure*}

\section{Star Forming Galaxies}

We fit star-forming galaxies with the same set of models (BC03, FSPS10, FSPS-C3K) with two different star formation histories: a model with constant star formation (CSF), and 
one with an exponentially declining SFR in the form of $\rm SFR(t) \sim exp(-t / \tau)$, with $\rm \tau = 1 Gyr$.

\subsection{Quality of fits}
\label{Quality of fits SFG}

Figure \ref{stacksSFG05} summarizes the best fits to the SFG sample at $0.5<z<1.0$ obtained with these combinations of models, letting the age $t$ vary. We again mask a 400$\rm \AA$ wide region around H$\alpha$ in the fit. According the $\chi_{red}^2$ statistics, the models assuming an exponentially declining SFH provide marginally beter fits than models assuming a CSF.

Figures \ref{stacksSFG10} and  \ref{stacksSFG15} show the best fits to the SFG samples at $1.0<z<1.5$ and $1.5<z<2.0$ obtained with the same combination of models. We mask 400$\rm \AA$ wide regions around
the expected strongest emission lines ([OIII], H$\alpha$). In the highest redshift stack, we see the biggest residual corresponding to the wavelength of the H$\gamma$ line ($\rm \lambda  = 4341 \AA$). For these redshifts, FSPS10 and FSPS-C3K model have a lower best-fit $\chi_{red}^2$ than that of BC03. However the models fit almost equally well.

Star-forming galaxies of these masses are in fact known to follow declining star-formation histories at redshifts lower than 1.5 (e.g. Pacifici et al. 2012).
As in the case of quiescent galaxies, different models have qualitatively and quantitatively different residuals. In this case however BC03 is the model with the smallest residuals, especially at the longest wavelengths ($\rm > 8500 \AA$).

\begin{figure}[!h!]
\centering
\includegraphics[width=8.8cm]{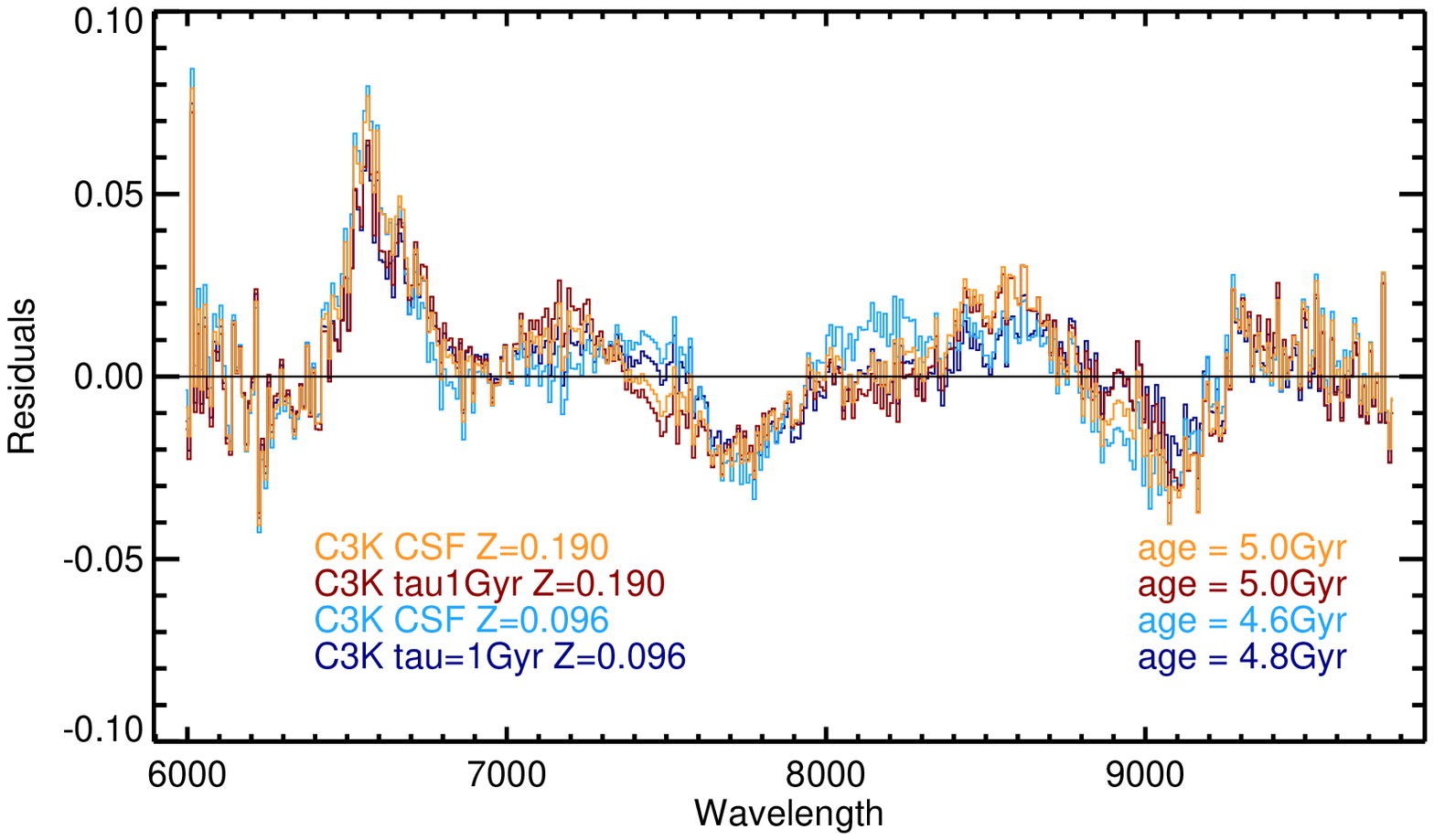}

\caption{For star-forming galaxies at redshift $0.5<z<1.0$, we compare residuals to the best-fits obtained by varying metallicity and star formation history.
The shape of residuals changes less than by changing the assumed stellar population model (compare with Figure \ref{stacksSFG05}). The maximum difference in the $\chi_{red}^2$ statistics for best fits with varying metallicity
is $\Delta \chi_{red}^2 = 0.2$, negligible in comparison to the difference obtained by varying the stellar population model ($\Delta \chi_{red}^2 = 5$).}
\label{stacksSFGresmet}
\end{figure}

The effect of varying metallicity is explored in Figure \ref{stacksSFGresmet} and compared to that of varying stellar population models.
We perform this test on star-forming galaxies at the lowest redshift bin ($0.5<z<1.0$), where the signal-to-noise ratio is high. Figure \ref{stacksSFGresmet} shows a comparison between residuals from best-fit FSPS-C3K models with solar metallicity models (Z=0.190) and models with half of the solar metallicity (Z=0.096), for a 1 Gyr $\tau$ model SFH, and a Constant SFR. We notice that the difference in ages is smaller than 10$\%$ and that residuals do not vary significantly. We conclude that the difference between different SPS models is greater than that obtained by using the same SPS code, with different metallicities and/or SFHs. 

%

\subsection{Determination of Ages}
\label{Determination of Ages SFG}

For star-forming galaxies at the lowest redshift we examine ($0.5<z<1.0$, Figure \ref{stacksSFG05}), the overall best-fit ($\chi_{red}^2 = 3.71$) is obtained with a young (0.6 Gyr of age) stellar population with the BC03 $\tau$ model. 
We also obtain a young age (0.2 Gyr) when assuming a constant star formation history for the same stellar population model.
The age determinations from FSPS10 and FSPS-C3K indicate instead an older age, from 1 to 5 Gyr (Figure \ref{stacksSFG05}).  We notice that for star-forming galaxies we cannot exclude any particular age range, since the stellar ages inferred from different models vary greatly.

At intermediate redshift ($1.0<z<1.5$, Figure \ref{stacksSFG10}), we obtain ages around 2-3 Gyrs with different models (the typical error on each inferred age for star-forming galaxies is 1 Gyr). 
At the highest redshifts ($1.5<z<2.0$, Figure \ref{stacksSFG15}), all best-fits (with different stellar population models and different star formation histories) converge to the lowest age value. 

For galaxies with active star formation, the intrinsic strength of features does not vary significantely with time since the spectra are dominated by light from young stars that been constantly forming. With the current signal-to-noise, we therefore cannot draw any conclusion on the ages of star-forming galaxies at the redshift under consideration.

\section{Discussion}

\subsection{Differences among SPSs}

\begin{figure*}[!t!h]
\centering
\includegraphics[width=15.2cm]{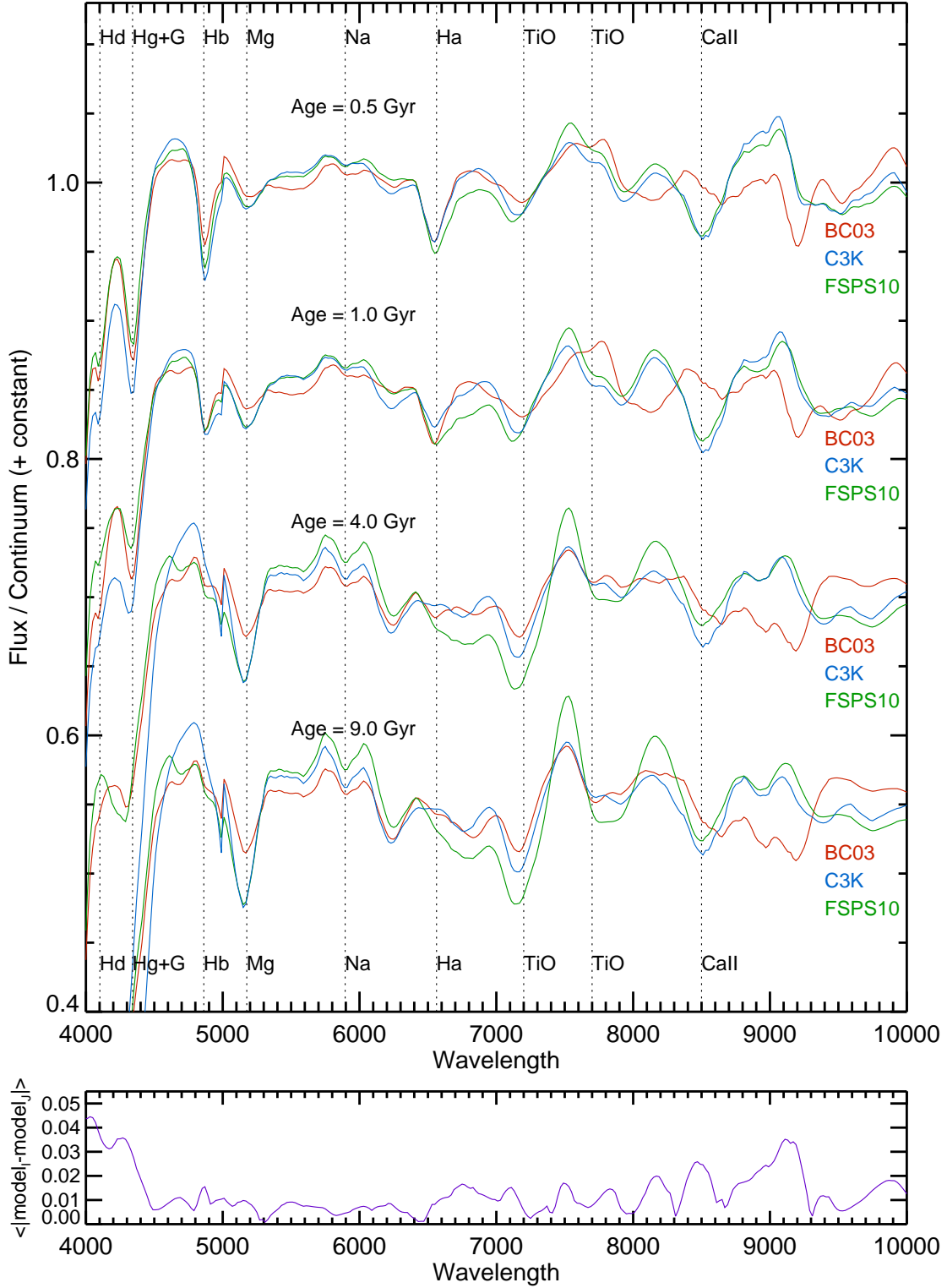}
\caption{Top: SSPs with solar metallicities from BC03 (red), FSPS10 (green), FSPS-C3K (blue). Models are convolved to 3D-HST resolution (see Section \ref{Methods}). Bottom: The purple line shows the average absolute difference between models at different wavelengths, for every combination of models of the same age: the difference among models is the biggest at the longest wavelengths and in the D4000 region.}
\label{different_models}
\end{figure*}

In order to investigate the origin of the qualitative difference in the best fits described in Section \ref{Quiescent Galaxies}, we compare model spectra from different SPS codes.
We show model SSPs from the BC03, FSPS10 and FSPS-C3K codes in Figure \ref{different_models}, for different ages. The strengths of the absorption lines vary among models, for every age. We notice in particular that at wavelengths higher than $\rm 7500 \AA$ different models predict different absorption bands at different wavelengths. We quantify the spread in the models at different wavelengths by computing the mean difference between every possible combination of models at the same age (Figure \ref{different_models}, bottom). This value is lower than 1\% at wavelength between $\rm \sim 4500 \AA$ and $\rm 6500 \AA$, and is larger otherwise. In particular the region with wavelengths greater than $\rm 8000 \AA$ has a large discrepancy between BC03 on one side, and FSPS10 and FSPS-C3K on the other.
This explains why determinations of ages from 3D-HST at higher redshifts are more stable between different models than those at lower redshift. For our lowest redshift sample ($0.5 < z < 1.0$) we observe the region of the spectrum where discrepancies among models are the largest, while at high redshift we observe rest-frame wavelengths where models are more similar to each others.

\subsection{Evolution of Ages}

\begin{figure*}[!t!h]
\centering
\includegraphics[width=8cm]{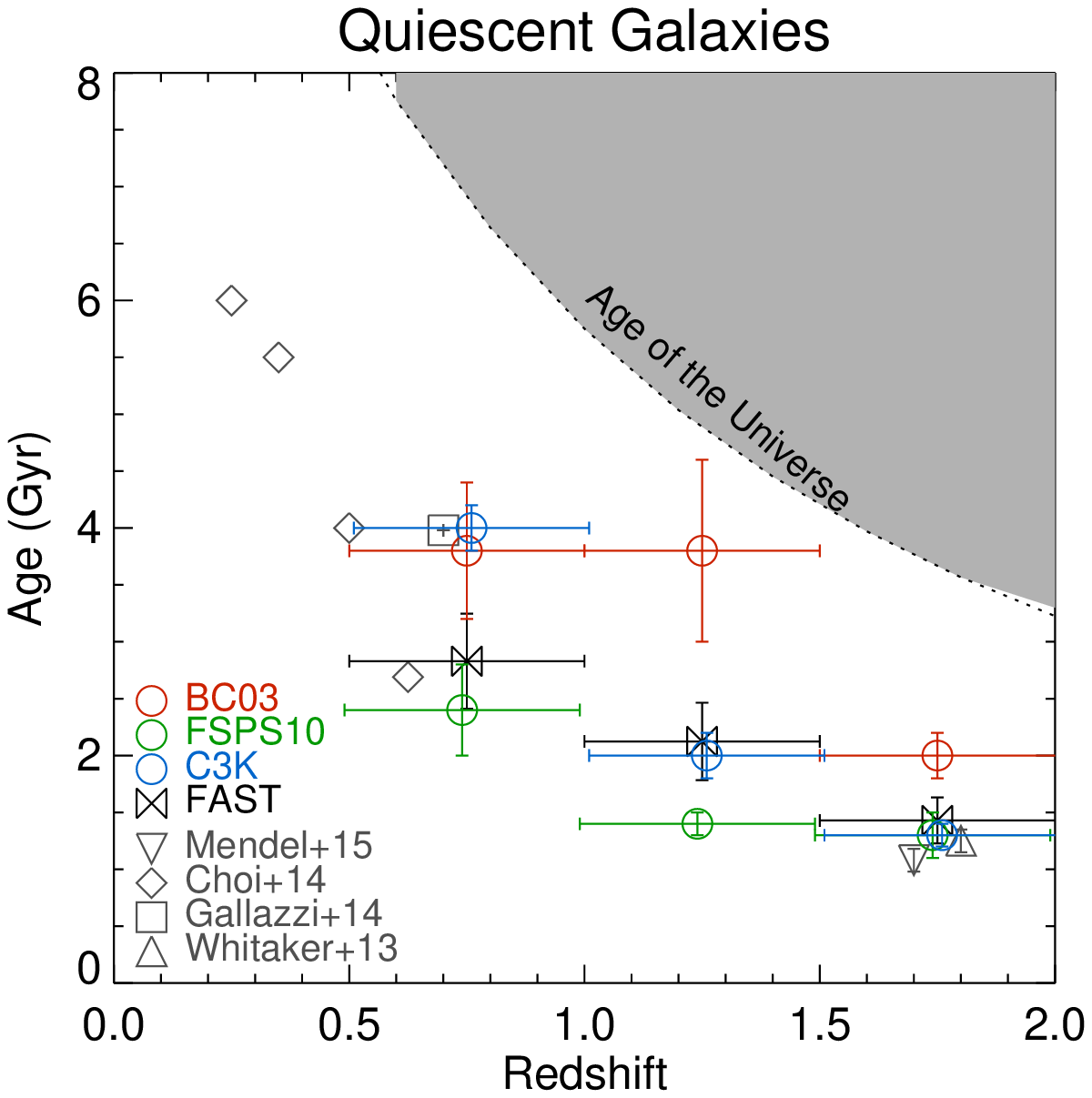}
\includegraphics[width=8cm]{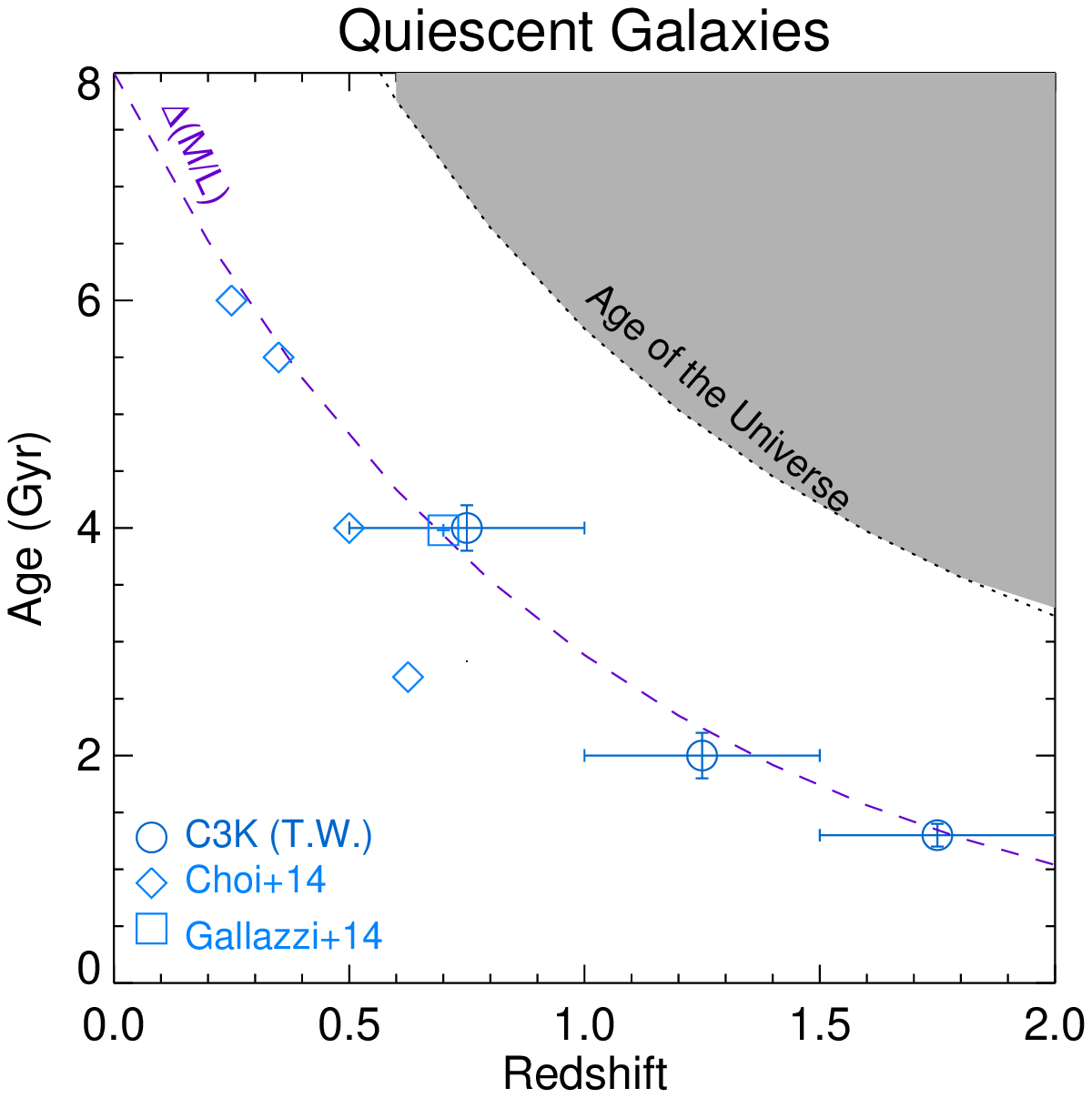}

\caption{Left: Evolution of ages of massive quiescent galaxies ($\rm log(M_{*} / M_{\odot}) > 10.8$) with redshift. Open circles represent values measured from 3D-HST. Different colors represent
different stellar population synthesis models (red: BC03, green: FSPS10, blue: FSPS-C3K) used for the determination of ages. As a comparison, values inferred from photometry with the FAST code (Kriek et al. 2009, black bow tie symbols) and from the literature selected in a similar mass range (grey symbols) are plotted.
Right: Comparison between the ages determined from 3D-HST spectra and the literature (blue) to the evolution of ages predicted from the evolution of the mass-to-light ratio inferred from the fundamental plane.
}
\label{evolutionages}
\end{figure*}

We investigate the evolution of ages of quiescent galaxies in a mass limited sample. In Fig. \ref{evolutionages} (left) we show the ages obtained by fitting 3D-HST stacks with different stellar population synthesis models. Even though the model dependent spread in ages is large, we observe that quiescent galaxies are younger at higher redshift and that, at each redshift, quiescent galaxies are not maximally old; instead their age is smaller than half of the age of the Universe at the same redshift.
We compare to data in a similar mass range by Gallazzi et al. (2014) at $z \sim 0.6$ and by Whitaker et al. (2013, who also uses spectra from 3D-HST) at  $1.4 < z < 2.2$, obtaining good agreement.
Also, at lower redshifts, Choi et al. (2014) find that quiescent galaxies at $0.2<z<0.7$ tend to be younger than half of the age of the Universe at those redshifts. The age value from the highest redshift mapped by Choi et al. (2014) is lower than the FSPS-C3K determination from this work, but similar to the determination obtained with the FSPS10 code, used by Choi et al. (2014) as well. We also note that the selection criteria and the rest-frame wavelength regime in Choi et al. (2014) are different from those in the present study. We test if the discrepancy between Choi et al. (2014) and our study can be caused by a selection bias, by repeating the spectral stacking and the fitting with the same sSFR selection of Choi et al. (2014). We find an age of 1.8 $\pm$ 0.6 Gyr, younger than our previous determination, in accordance to Choi et al. (2014).

Fairly young ages of quiescent galaxies are expected at $z \sim 1$ from the mass build-up of galaxies: a subset of the quiescent galaxies were likely still star forming at slightly higher redshifts. 
In addition, quiescent galaxies may grow by merging with lower mass star-forming galaxies. The processes involved are clearly complex but a simple estimate of the expected ages of z$\sim$1 quiescent galaxies can be obtained from comparisons of comoving stellar mass densities of quiescent galaxies and star-forming galaxies.  

Muzzin et al. (2014) find a stellar mass density of quiescent galaxies at $z \sim 1$ of log($\rho_{*}$ $\rm [ M_{\sun} Mpc ^{-3}] ) = 7.8$ (their Fig. 8). This density was reached at a redshift of $z \sim 2.1$ for all galaxies (that is, quiescent and star-forming galaxies). If the oldest stars at $z\sim1$ are in the quiescent galaxiy population, these stars finished forming at a redshift of $z \sim 2.1$. The time difference between $z=2.1$ and $z=1$ is 2.6 Gyr. This is then the youngest age of this population, while the median age is defined by the redshift at which half of the mass was formed ($z \sim 2.4$). The resulting median age is 3 Gyr, similar to that we observe in this study (Fig. \ref{evolutionages}, right).  This is strictly an upper limit to the age since z=1 quiescent galaxies can contain stars that formed later. On the other hand this argument is strictly valid for the full population of quiescent galaxies, and lower mass quiescent galaxies might be somewhat younger.  It is remarkable that this simple argument gives an age which agrees well with our general results.

\subsubsection{Comparison with photometry}

Ages of galaxies can also be inferred from photometry only. In 3D-HST stellar masses, star-formation rates, ages and dust extinction are estimated with the FAST code (Kriek et al. 2009), assuming exponentially declining star formation histories with
a minimum e-folding time of $\rm log_{10}(\tau/yr) = 7$, a minimum age of 40 Myr, $0< A_{V} <4$ mag and the Calzetti et al. (2000) dust attenuation law (see Skelton et al. 2014).
The output from the FAST code is an age defined as the time since the onset of SF, that is not necessary equivalent to a light-weighted age.

For each galaxy we therefore compute a light-weighted age ($\rm t_{lum}$) following the definition:

$$\rm t_{lum} = \frac{\sum_{i} SFR(t_{i}) \times V_{SSP}(t-t_{i}) \times (t-t_{i}) \times \Delta_{t}}{\sum_{i} SFR(t_{i}) \times V_{SSP}(t-t_{i}) \times \Delta_{t}}$$
where:

\begin{itemize}
  \item $t$ is the time of observation (equivalent to the age of FAST)
  \item $\rm SFR(t_{i})$ is the star formation rate at time $t_{i}$. In the case of the FAST fits $\rm SFR(t_{i})$ has the functional shape of a $\tau$ model $\rm SFR(t_{i}) \sim exp(-t_{i} / \tau)$ 
  \item $V_{SSP}(t-t_{i})$ is the V-band flux of 1 $M_{\odot}$ element formed at $t_{i}$ and observed at time $t$.
  \item $\Delta_{t}$ is the time-step we divide the SFH in (we use  $\Delta_{t} = 50$ Myr).

\end{itemize} 
We notice that for an SSP $\rm t_{lum}(t) = t$.

\begin{figure*}[!t!h]
\centering
\includegraphics[width=7cm]{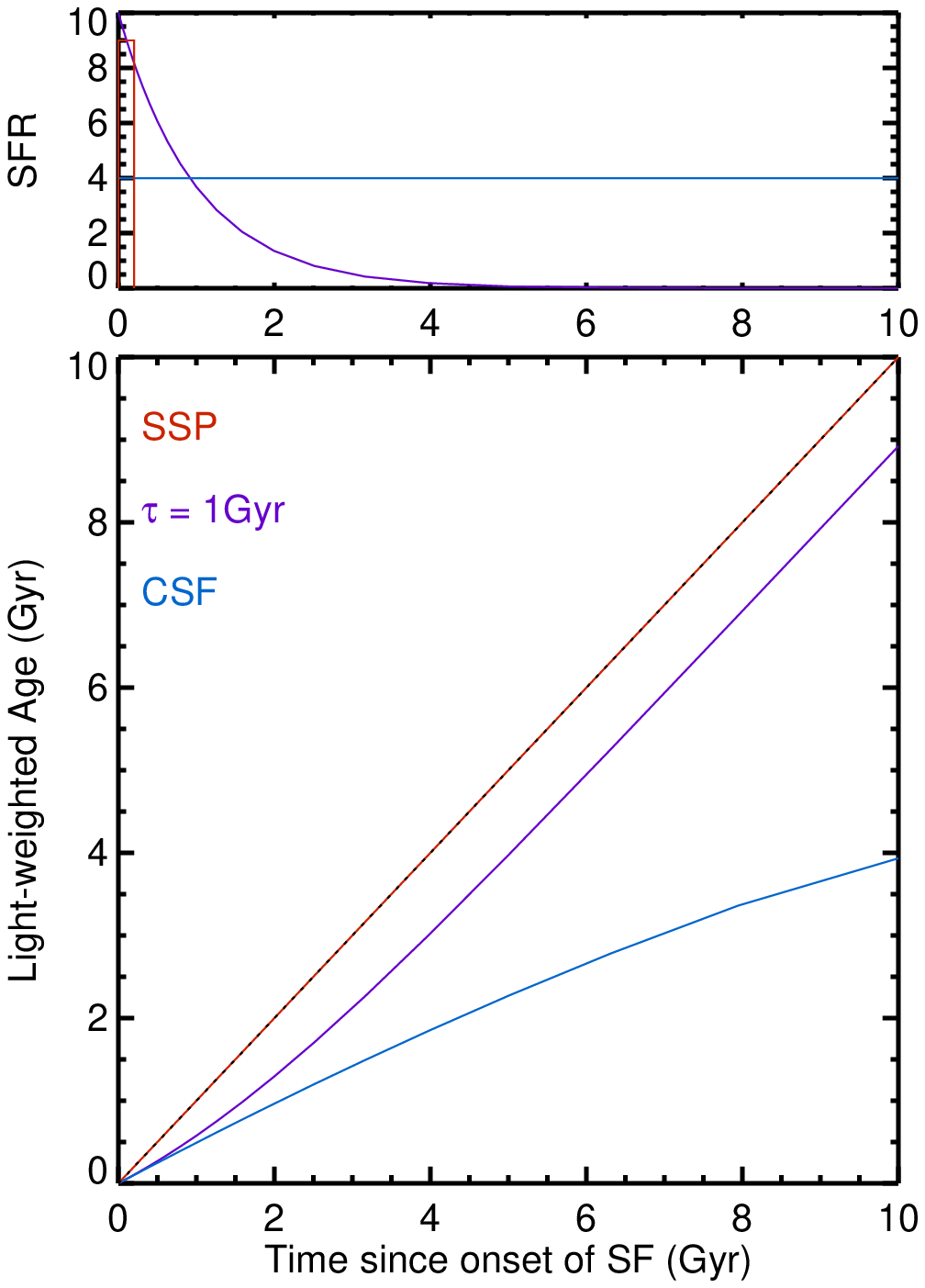}
\includegraphics[width=7cm]{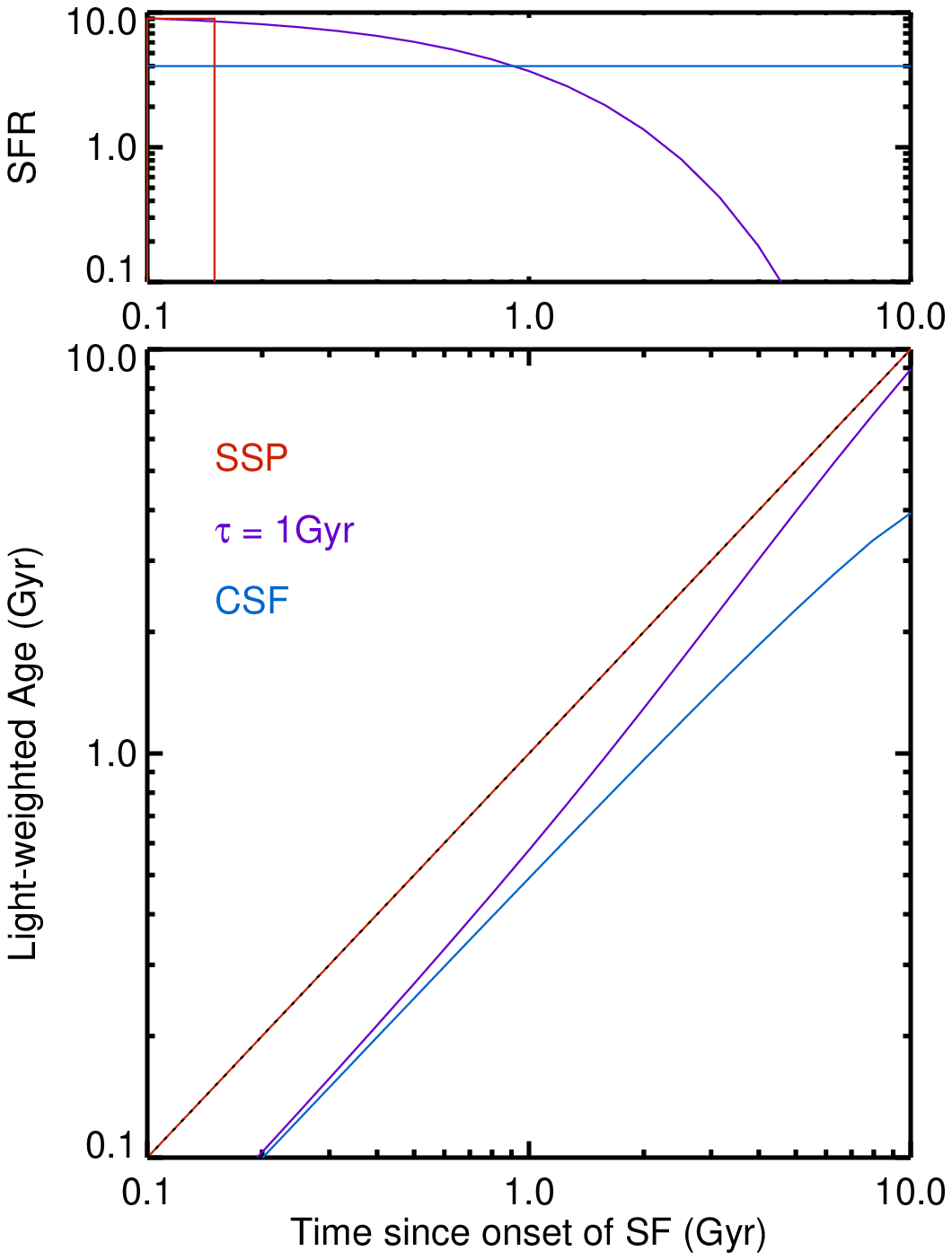}

\caption{Light-weighted ages ($\rm t_{lum}$) at different times from the onset of star-formation, in linear (left) and logarithmic scale (right). Three different SFH are shown: single stellar population (SSP, red), constant star formation (CSF, blue), and an exponentially declining $\tau$-model with $\tau = 1 Gyr$. For a SSP the light weighted age corresponds to the time from the burst. For a CSF $\rm t_{lum}$ is always lower than the time from the onset of star-formation, with approximately $\rm t_{lum} \sim t / 2$. A $\tau$ model has an intermediate behaviour, being asymptotically similar to the CSF at $t \rightarrow 0$ and to the 1-1 slope at later times.}
\label{lwamodels}
\end{figure*}

Figure \ref{lwamodels} shows the relation between the time from the onset of star-formation $t$ and $\rm t_{lum}$ for a range of models: an SSP, a CSF model, and an exponentially declining model with $\rm \tau = 1 Gyr$. As expected,$\rm t_{lum}$  for an SSP (red line) is well approximated by $t$. For a CSF (blue line) $\rm t_{lum}$ is always smaller than $t$, with a increasing difference at later times. This effect is naturally explained by the fact that younger stars are brighter than older stars: $V_{SSP}$ peaks at 10 Myr, and declines afterwards (in other words the mass-to-light ratio $M/L_{V}$ increases for older stellar populations, see among others Bruzual \& Charlot 2003, Fig 1 to 5).
The $\tau$ model (purple line) has an intermediate behavior, being similar to the CSF for $t \rightarrow 0$, and parallel to the SSP for large $t$. We note that these relations are well known (see, e.g., Appendix A of van Dokkum et al. 1998).

For each quiescent galaxy in the sample, we infer $\rm t_{lum}$ from the best fit to its photometry, and find the average value in the three redshift bins $0.5<z<1.0$, $1.0<z<1.5$ and $1.5<z<2.0$. Figure \ref{evolutionages} (left) shows how the quantity compares to the ages measured from the spectra (Section \ref{Quiescent Galaxies}) with different sets of models.
For each redshift bin ages derived from photometry tend to be comparable to the lowest values obtained with the spectral fitting.

\subsubsection{Comparison with fundamental plane studies}

Another technique commonly used to constrain ages of high redshift quiescent galaxies is provided by the fundamental plane (Djorgovski \& Davis 1987, hereafter: FP).
In particular, the FP is a model-independent tool for measuring the mass to light ratio (M/L).
The offset between the M/L of high redshift galaxies and that of local galaxies can therefore be used to estimate the age of their
stellar populations (Franx 1993, van Dokkum \& Franx 1996, van der Wel et al. 2004, Treu et al. 2005).
Since the luminosity of an SSP evolves with time as $\rm L \sim t^k$, (with the parameter k derived from stellar population models),
the relation between the evolution of M/L and light-weighted ages can therefore be approximated as
$\rm \Delta ln (M/L) \sim -k \Delta ln(t)$.
Measurements of the evolution of $M/L$ up to $z\sim 1$ agree with values $\rm \Delta ln (M/L_{B}) \sim -1 \times z$
(van Dokkum \& Stanford 2003, Wuyts et al. 2004, Holden et al. 2005).
Given a value of k=-0.98 (from BC03 models in the B band, with a Chabrier IMF), 
we derive the following relation between the local light-weighted ages and those at high redshift: 
$t_{lum}(z) = t_{lum}(0) \times e^{-z/0.98}$.

The observed evolution of the $M/L$ predicts that the ages of quiescent galaxies at $z \sim 1$ are 2.8 times younger than those at $z=0$, and 4 times younger at $z\sim1.5$.
Figure \ref{evolutionages} shows the age evolution predicted from M/L measurements. We use 8 Gyr as the age of galaxies at $z=0$, as measured from SDSS spectra by Gallazzi et al. (2006).
The agreement between the FP prediction and measurements from 3D-HST spectra is excellent. Only the spectral measurement with the BC03 models at $z\sim 1.25$ and the one with the FSPS10 models at $z\sim 0.75$
significantly deviate from the FP prediction.
\subsection{H$\alpha$ in quiescent galaxies}
\label{Halpha in quiescent galaxies}

At redshifts lower than 1.5 we can quantify the H$\alpha$ emission\footnote{Given the WFC3 grism resolution H$\alpha$ and [NII] are inevitably blended, therefore we will refer to measurements of H$\alpha$+[NII]} in quiescent galaxies from the residuals to the best fits (Figure \ref{bestfitsQG05} and \ref{bestfitsQG10}).  
We subtract the best-fit model with the FSPS-C3K SPS from the stacks, and fit residuals with a Gaussian centered at the H$\alpha$ wavelength ($\lambda =$6563$\rm \AA$).
Since the stacks are continuum subtracted, this is effectively a direct measurement of EW(H$\alpha$+[NII]).
At the lowest redshifts ($0.5<z<1.0$), we do not obtain a significant detection, with $\rm EW(H\alpha+[NII])=0.5 \pm 0.3 \AA$, 
while at $1.0<z<1.5$ we robustly detect the emission line, measuring EW(H$\alpha$+[NII]) = $\rm 5.5 \pm 0.8 \AA$.    
In the same mass/redshift regime typical star-forming galaxies have EW(H$\alpha$+[NII]) $\rm \sim 60 \AA$ (Fumagalli et al. 2012). This shows that, assuming no dust absorption, the H$\alpha$ emission in quiescent galaxies is quenched by a factor of $\sim 10$.

To estimate the H$\alpha$ fluxes, we multiply the EW(H$\alpha$+[NII]) by the median continuum flux of galaxies in the stack, and assume a 0.25 ratio for [NII]/(H$\alpha$+[NII]). We finally estimate SFR(H$\alpha$)
with the Kennicutt (1998) relation, obtaining that quiescent galaxies have $\rm SFR(H\alpha) = 0.46 \pm 0.06 M_{\odot}/yr$ at $1.0<z<1.5$ and $0.10 \pm 0.05 M_{\odot}/yr$ at $0.5<z<1.0$, assuming no dust absorption. 

In Fumagalli et al. (2014) we reported for quiescent galaxies higher SFR inferred from mid-infrared emissions, up to $\rm 3.7 \pm 0.7 M_{\odot}/yr$ at $1.1<z<1.5$.  Since the sample in Fumagalli et al. (2014) has a slightly different selection ($1.1 < z < 1.5$, log$(M^*/M_{\odot}) >$  10.3), we test if the discrepancy between H$\alpha$ and IR inferred SFRs can be caused by a selection bias, by repeating the spectral stacking and the H$\alpha$ measurement in the same redshift / mass window of Fumagalli et al. (2014). We obtain
SFR(H$\alpha$)$\rm = 0.32 \pm 0.07 M_{\odot}/yr$, which would require a significant dust extinction for the H$\alpha$ line ($A_{H\alpha} \sim$  2.6) to be reconciled with the SFR(IR) measurement. This extinction value is similar to the A(H$\alpha$) measured for star-forming galaxies of similar masses at both redshift z~1.5 (Sobral et al. 2012, Kashino et al. 2013, Price et al. 2014, Fumagalli et al. 2015) and in the local Universe (Garn \& Best 2010) with no significative redshift dependence.

A variety of studies (Fumagalli et al. 2014, Utomo et al. 2014, Hayward et al. 2014) suggests that SFR inferred from IR are overestimated because of the contribution of dust heating by old stars and/or TP-AGB stars to the MIR fluxes. SFRs measured from H$\alpha$ are instead contaminated by potential AGN or LINER emission and affected by dust extinction. Our combined multiwavelength findings agree however in indicating that SFRs of quiescent galaxies are very low, they are negligible in comparison to those of star-forming galaxies at the same redshift, and they are potentially consistent with 0. 


\section{Conclusions}

We select massive galaxies from the 3D-HST survey and divide them into quiescent and star-forming according to their rest-frame optical and near-infrared colors.
We stack their low-resolution spectra from 3D-HST in three redshift bins, and fit them with models from three stellar population synthesis codes, in order to infer the mean stellar ages of the sample.
 
For quiescent galaxies, we show that the new FSPS-C3K models provides more accurate fits to the data. Other codes do not reproduce the observed features at the reddest optical wavelengths. 

For star-forming galaxies, we are not able to put significant constraints on the stellar ages of the samples.

Even though we infer different stellar ages from different models, stellar ages of quiescent galaxies appear to be overall younger than half of the age of the Universe, confirming the trends found at lower redshift by Choi et al. (2014) and Gallazzi et al. (2014). The evolution of stellar ages is moreover in accordance with the expected evolution from fundamental plane studies.

\acknowledgments
We thank the referee for the constructive comments that improved the quality of the paper. We thank Charlie Conroy for the support through the development of this paper, for the useful discussions and for sharing his stellar population synthesis models. We thank Jesse van de Sande and Adam Muzzin for the useful discussions. 
We acknowledge funding from ERC grant HIGHZ no. 227749 and a NWO Spinoza grant. This work is based on observations taken by the 3D-HST Treasury Program (GO 12177 and 12328) with the NASA/ESA HST, which is operated by the Association of Universities for Research in Astronomy, Inc., under NASA contract NAS5-26555. Support from NASA/STScI through program GO-12177 is gratefully acknowledged. KEW gratefully acknowledges support by NASA through Hubble Fellowship grant \# HF2-51368 awarded by the Space Telescope Science Institute, which is operated by the Association of Universities for Research in Astronomy, Inc., for NASA, under contract NAS 5-26555.

\end{document}